\documentclass[11pt]{article}

\usepackage{times}
\usepackage{amsmath}
\usepackage{amsfonts}
\usepackage{latexsym}
\usepackage{fullpage}

\begin{document}
\newcommand{\mytheoremcounter}{section}
\newcommand{\cheapHack}{ }
\newcommand{\True}{\mathrm{true}}
\newcommand{\False}{\mathrm{false}}
\newcommand{\Sarg}[2]{{\scriptscriptstyle{\stackrel{#1}{#2}}}}
\newcommand{\ChgBar}{\marginpar{$\spadesuit$}}
\newcommand{\fixme}[1]{$\spadesuit$\marginpar{\tiny$\spadesuit$#1}}
\newcommand{\Xhalf}{X_\frac{1}{2}}
\newcommand{\Shalf}{\cS_{\frac{n}{2},n}}
\newcommand{\chih}{\chi_n}
\newcommand{\notchih}{\overline{\chi_n}}
\newcommand{\upsh}{\upsilon_n}
\newcommand{\notupsh}{\overline{\upsilon_n}}
\newcommand{\singlespacing}{\baselineskip 1em}
\newcommand{\onehalfspacing}{\baselineskip 1.25em}
\newcommand{\doublespacing}{\baselineskip 1.75em}
\newcommand{\truedoublespacing}{\baselineskip 2em}
\newcommand{\normalspacing}{\singlespacing}
\newcommand{\Paccept}[1]{{\Pr[#1 = \mathrm{accept}]}}
\newcommand{\Maj}{\mathit{Maj}}
\newcommand{\Nand}{\uparrow}
\newcommand{\Nor}{\downarrow}
\newcommand{\myem}[1]{{\bf #1}}
\newcommand{\myit}[1]{{\it #1}}
\newcommand{\reccr}[1]{\overline{\nu}(#1)}
\newcommand{\regcr}[1]{\nu(#1)}
\newcommand{\LOGDCFLclass}{\mathbf{LOGDCFL}}
\newcommand{\NLclass}{\mathbf{NL}}
\newcommand{\ACclass}{\mathbf{AC}}
\newcommand{\NCclass}{\mathbf{NC}}
\newcommand{\SCclass}{\mathbf{SC}}
\newcommand{\RCclass}{\mathbf{RC}}
\newcommand{\coNLclass}{\mathbf{co\!-\!NL}}
\newcommand{\Lclass}{\mathbf{L}}
\newcommand{\Lpoly}{\mathbf{L}_{/\mathrm{poly}}}
\newcommand{\Pclass}{\mathbf{P}}
\newcommand{\BPclass}{\mathbf{P}_{BP}}
\newcommand{\BPwidth}[1]{\mathbf{P}_{BP}^{#1}}
\newcommand{\NPclass}{\mathbf{NP}}
\newcommand{\coNPclass}{\mathbf{coNP}}
\newcommand{\PPclass}{\mathbf{PP}}
\newcommand{\BPPclass}{\mathbf{BPP}}
\newcommand{\ZPPclass}{\mathbf{ZPP}}
\newcommand{\RPclass}{\mathbf{RP}}
\newcommand{\coRPclass}{\mathbf{co\!-\!RP}}
\newcommand{\SigmaP}[1]{\mathbf{\Sigma_{#1}P}}
\newcommand{\PHclass}{\mathbf{PH}}
\newcommand{\PSPACE}{\mathbf{PSPACE}}
\newcommand{\RSPACE}{\mathbf{RSPACE}}
\newcommand{\DSPACE}{\mathbf{DSPACE}}
\newcommand{\imin}{{\mathit{min}}}
\newcommand{\imax}{{\mathit{max}}}
\newcommand{\Supp}{{\mathit{supp}}}
\newcommand{\gbf}{{\mathbf{g}}}
\newcommand{\zbf}{{\mathbf{z}}}
\newcommand{\wbf}{{\mathbf{w}}}
\newcommand{\vbf}{{\mathbf{v}}}
\newcommand{\xbf}{{\mathbf{v}}}
\newcommand{\onebf}{{\mathbf{1}}}
\newcommand{\zerobf}{{\mathbf{0}}}
\newcommand{\uvec}{\vec{u}}
\newcommand{\vvec}{\vec{v}}
\newcommand{\xvec}{\vec{x}}
\newcommand{\Sigmap}{\Sigma^\prime}
\newcommand{\alphap}{\alpha^\prime}
\newcommand{\betap}{\beta^\prime}
\newcommand{\betapp}{\beta^{\prime\prime}}
\newcommand{\gammap}{\gamma^\prime}
\newcommand{\mup}{\mu^\prime}
\newcommand{\mupp}{\mu^{\prime\prime}}
\newcommand{\nubar}{{\overline{\nu}}}
\newcommand{\nubars}{\nubar^*}
\newcommand{\nus}{\nu^*}
\newcommand{\betas}{\beta^*}
\newcommand{\deltas}{\delta^*}
\newcommand{\deltap}{\delta^\prime}
\newcommand{\deltapp}{\delta^{\prime\prime}}
\newcommand{\lambdabar}{\overline{\lambda}}
\newcommand{\lambdap}{\lambda^\prime}
\newcommand{\lambdapp}{\lambda^{\prime\prime}}
\newcommand{\pip}{\pi^\prime}
\newcommand{\pipp}{{\pi^{\prime\prime}}}
\newcommand{\Psip}{\Psi^\prime}
\newcommand{\psip}{\psi^\prime}
\newcommand{\phip}{\phi^\prime}
\newcommand{\sigmap}{\sigma^\prime}
\newcommand{\sigmat}{\tilde{\sigma}}
\newcommand{\taup}{\tau^\prime}
\newcommand{\etap}{\eta^\prime}
\newcommand{\etapp}{{\eta^{\prime\prime}}}
\newcommand{\epsilonp}{{\epsilon^\prime}}
\newcommand{\epsilonpp}{\epsilon^{\prime\prime}}
\newcommand{\epsilonppp}{\epsilon^{\prime\prime\prime}}
\newcommand{\ap}{a^\prime}
\newcommand{\bp}{b^\prime}
\newcommand{\cp}{c^\prime}
\newcommand{\up}{u^\prime}
\newcommand{\fp}{f^\prime}
\newcommand{\ft}{\tilde{f}}
\newcommand{\ep}{e^\prime}
\newcommand{\gp}{g^\prime}
\newcommand{\gtd}{\tilde{g}}
\newcommand{\ip}{i^\prime}
\newcommand{\jp}{j^\prime}
\newcommand{\mpr}{m^\prime}
\newcommand{\np}{n^\prime}
\newcommand{\op}{o^\prime}
\newcommand{\qp}{q^\prime}
\newcommand{\rp}{r^\prime}
\newcommand{\gpp}{g^{\prime\prime}}
\newcommand{\ipp}{i^{\prime\prime}}
\newcommand{\jpp}{j^{\prime\prime}}
\newcommand{\epp}{e^{\prime\prime}}
\newcommand{\qpp}{q^{\prime\prime}}
\newcommand{\rpp}{r^{\prime\prime}}
\newcommand{\vp}{v^\prime}
\newcommand{\ypp}{y^{\prime\prime}}
\newcommand{\xpp}{x^{\prime\prime}}
\newcommand{\xt}{\tilde{x}}
\newcommand{\zp}{z^\prime}
\newcommand{\hp}{{h^\prime}}
\newcommand{\lp}{{l^\prime}}
\newcommand{\zpp}{{z^{\prime\prime}}}
\newcommand{\kp}{{k^\prime}}
\newcommand{\Dp}{{D^\prime}}
\newcommand{\Pp}{P^\prime}
\newcommand{\Ppp}{P^{\prime\prime}}
\newcommand{\Qp}{Q^\prime}
\newcommand{\Qpp}{Q^{\prime\prime}}
\newcommand{\Spr}{S^\prime}
\newcommand{\Tp}{T^\prime}
\newcommand{\Ep}{E^\prime}
\newcommand{\Dbar}{\overline{D}}
\newcommand{\Lp}{L^\prime}
\newcommand{\Lpp}{L^{\prime\prime}}
\newcommand{\Lbar}{\overline{L}}
\newcommand{\Lhat}{\widehat{L}}
\newcommand{\Ltilde}{\tilde{L}}
\newcommand{\Lcap}{{L^\cap}}
\newcommand{\Lcup}{{L^\cup}}
\newcommand{\Lpbar}{\overline{\Lp}}
\newcommand{\Mp}{M^\prime}
\newcommand{\Mpp}{M^{\prime\prime}}
\newcommand{\Mbar}{\overline{M}}
\newcommand{\Np}{N^\prime}
\newcommand{\Npp}{N^{\prime\prime}}
\newcommand{\Rp}{R^\prime}
\newcommand{\xbar}{\bar{x}}
\newcommand{\xp}{x^\prime}
\newcommand{\yp}{y^\prime}
\newcommand{\Uhat}{\widehat{U}}
\newcommand{\Up}{U^\prime}
\newcommand{\Upp}{U^{\prime\prime}}
\newcommand{\Vp}{V^\prime}
\newcommand{\Vhat}{\widehat{V}}
\newcommand{\Vbar}{\overline{V}}
\newcommand{\Ap}{{A^\prime}}
\newcommand{\App}{{A^{\prime\prime}}}
\newcommand{\Cp}{{C^\prime}}
\newcommand{\Cpp}{C^{\prime\prime}}
\newcommand{\Fp}{F^\prime}
\newcommand{\Gp}{G^\prime}
\newcommand{\Gtilde}{\tilde{G}}
\newcommand{\Fpp}{F^{\prime\prime}}
\newcommand{\Zf}{{\mathbb{Z}}}
\newcommand{\Qf}{{\mathbb{Q}}}
\newcommand{\Rf}{{\mathbb{R}}}
\newcommand{\Cf}{{\mathbb{C}}}
\providecommand{\mathbb}[1]{\Bbb{#1}}
\newcommand{\qacc}{q^{acc}}
\newcommand{\qrej}{q^{rej}}
\newcommand{\qnon}{q^{non}}
\newcommand{\Qadd}{Q_{add}}
\newcommand{\Qacc}{Q_{acc}}
\newcommand{\Qrej}{Q_{rej}}
\newcommand{\Qnon}{Q_{non}}
\newcommand{\Qhalt}{Q_{halt}}
\newcommand{\Qjunk}{Q_{junk}}
\newcommand{\Qaccp}{{Q_{acc}^\prime}}
\newcommand{\Qrejp}{{Q_{rej}^\prime}}
\newcommand{\Qnonp}{{Q_{non}^\prime}}
\newcommand{\Qhaltp}{{Q_{halt}^\prime}}
\newcommand{\Qjunkp}{{Q_{junk}^\prime}}
\newcommand{\Qaccpp}{{Q_{acc}^{\prime\prime}}}
\newcommand{\Qrejpp}{{Q_{rej}^{\prime\prime}}}
\newcommand{\Qnonpp}{{Q_{non}^{\prime\prime}}}
\newcommand{\Qjunkpp}{{Q_{junk}^{\prime\prime}}}
\newcommand{\Cacc}{C_{acc}}
\newcommand{\Crej}{C_{rej}}
\newcommand{\Cnon}{C_{non}}
\newcommand{\Eacc}{E_{acc}}
\newcommand{\Erej}{E_{rej}}
\newcommand{\Enon}{E_{non}}
\def\cent{{\hbox{\rm\rlap/c}}}
\newcommand{\centp}{{\cent}^\prime}
\newcommand{\Bra}[1]{{\langle{#1}|}}
\newcommand{\Ket}[1]{{|{#1}\rangle}}
\newcommand{\BraKet}[2]{{\langle{#1}|{#2}\rangle}}
\newcommand{\iprod}[2]{{\langle{#1},{#2}\rangle}}
\newtheorem{theorem}{{\bf Theorem}}[\mytheoremcounter]
\newtheorem{lemma}[theorem]{{\bf Lemma}}
\newtheorem{claim}[theorem]{{\bf Claim}}
\newtheorem{example}[theorem]{{\bf Example}}
\newtheorem{question}[theorem]{{\bf Question}}
\newtheorem{answer}[theorem]{{\bf Answer}}
\newtheorem{conjecture}[theorem]{{\bf Conjecture}}
\newtheorem{proposition}[theorem]{{\bf Proposition}}
\newtheorem{corollary}[theorem]{{\bf Corollary}}
\newtheorem{fact}[theorem]{{\bf Fact}}
\newtheorem{definition}[theorem]{{\bf Definition}}
\newtheorem{remark}[theorem]{{\bf Remark}}
\newtheorem{thoughts}[theorem]{{\bf Thoughts}}
\newenvironment{proof}{ \begin{trivlist} 
                        \item \vspace{-\topsep} \noindent{\bf Proof:}\ }
                      {\rule{5pt}{5pt}\end{trivlist}}
\newcommand{\Case}[2]{\noindent{\bf Case #1:}#2}
\newcommand{\Subcase}[2]{\noindent{\bf Subcase #1:}#2}
\newcommand{\Half}{\frac{1}{2}}
\newcommand{\RtHalf}{\frac{1}{\sqrt{2}}}
\newcommand{\cA}{{\mathcal{A}}}
\newcommand{\cC}{{\mathcal{C}}}
\newcommand{\cE}{{\mathcal{E}}}
\newcommand{\cF}{{\mathcal{F}}}
\newcommand{\cH}{{\mathcal{H}}}
\newcommand{\cI}{{\mathcal{I}}}
\newcommand{\cK}{{\mathcal{K}}}
\newcommand{\cL}{{\mathcal{L}}}
\newcommand{\cM}{{\mathcal{M}}}
\newcommand{\cO}{{\mathcal{O}}}
\newcommand{\cP}{{\mathcal{P}}}
\newcommand{\cR}{{\mathcal{R}}}
\newcommand{\cS}{{\mathcal{S}}}
\newcommand{\cU}{{\mathcal{U}}}
\newcommand{\Span}{{\mathit{Span}}}
\newcommand{\Ch}[2]{{#1 \choose #2}}
\newcommand{\Ul}[1]{{\underline{#1}}}
\newcommand{\Floor}[1]{{\lfloor #1 \rfloor}}
\newcommand{\ignore}[1]{}
\newcommand{\noignore}[1]{#1}

\newcommand{\RMO}{\mathbf{RMO}}
\newcommand{\UMO}{\mathbf{UMO}}
\newcommand{\RMOe}{\mathbf{RMO}_\epsilon}
\newcommand{\RMM}{\mathbf{RMM}}
\newcommand{\UMM}{\mathbf{UMM}}
\newcommand{\RMMe}{\mathbf{RMM}_\epsilon}

\newcommand{\MOQFA}{\mathbf{MOQFA}}
\newcommand{\MOQFAe}{\mathbf{MOQFA}_\epsilon}
\newcommand{\MMQFA}{\mathbf{MMQFA}}
\newcommand{\MMQFAe}{\mathbf{MMQFA}_\epsilon}
\newcommand{\GQFA}{\mathbf{GQFA}}
\newcommand{\GQFAe}{\mathbf{GQFA}_\epsilon}

\newcommand{\REG}{\mathbf{REG}}
\newcommand{\PFA}{\mathbf{PFA}}
\newcommand{\PFAe}{\mathbf{PFA}_\epsilon}
\newcommand{\GFA}{\mathbf{GFA}}

% Code environment
\newcommand{\Foreach}[2]{\\{\bf\tt{for\ each}} $#1$ {\bf\tt{do}}\+ #2
\- \\ {\bf\tt{rof}}}
\newcommand{\Forloop}[2]{\\{\bf\tt{for}} $#1$ {\bf\tt{do}}\+ #2
\- \\ {\bf\tt{rof}}}
\newcommand{\Ifthen}[2]{\\{\bf\tt{if}} $#1$ {\bf\tt{then}}\+ #2
\- \\ {\bf\tt{fi}}}
\newcommand{\Ifelse}[3]{\\{\bf\tt{if}} $#1$ {\bf\tt{then}}\+ #2
\- \\ {\bf\tt{else}}\+ #3 \- \\ {\bf\tt{fi}}}
\newcommand{\Stmt}[1]{\\$#1$;}
\newcommand{\StartStmt}[1]{\+\kill$#1$;}
\newenvironment{pseudocode}{\begin{tabbing} 
\ \ \ \ \=\ \ \ \ \=\ \ \ \ \=\ \ \ \ \=\ \ \ \ \=\ \ \ \ \=\ \ \ \ \=\
\ \ \ \= } {\end{tabbing}}

% Moving proofs around
% #1 = proof \name   #2 = appendix ref    #3 = proof
\providecommand{\SaveProof}[3]{#3}
% #1 = proof \name   #2 = appendix ref    #3 = proof    #4 = sketch
\providecommand{\SketchProof}[4]{#3}
% #1 = proof \name   #2 = appendix label  #3 = title
\providecommand{\AppendixProof}[3]{}

\newcommand{\MoveProofsToAppendix}{\include{movemacs}}

% Short equation separator
\newcommand{\ShortSep}{\\ & &}
\newcommand{\LongSep}{}
\providecommand{\DefSep}{\LongSep}
\newcommand{\UseShortSep}{\renewcommand{\DefSep}{\ShortSep}}

% select abstract mechanism
\newcommand{\UseAbstract}[2]{#1}
\newcommand{\UseOtherAbstract}{\include{absselect}}

% EVIL stuff to make it fit for 10 page limit (two lines per page extra)
\newcommand{\StretchPage}{ \addtolength{\textheight}{0.05\textheight}
                           \addtolength{\topmargin}{-0.03\textheight}
                         }

% figure macro

\newcommand{\DoFigure}[4]{
                          \begin{figure}[ht]
                            \begin{center}
                              \ \includegraphics[scale=#2]{#1}\ 
                            \end{center}
                            \caption{#3\label{#4}}
                          \end{figure}
                         }

\newcommand{\DoBiFigure}[5]{
                          \begin{figure}[ht]
                            \begin{center}
                              \mbox{\ \includegraphics[scale=#3]{#1}\ 
                                    \hspace{1.0in}
                                    \ \includegraphics[scale=#3]{#2}\ }
                            \end{center}
                            \caption{#4\label{#5}}
                          \end{figure}
                         }

\newcommand{\DoDiFigure}[8]{ 
                          \begin{figure}[ht]
                            \begin{center}
                              \begin{minipage}[b]{0.35\linewidth}
                                \begin{center}
                                  \includegraphics[scale=#2]{#1}
                                \end{center}
                                \caption{#3\label{#4}}
                              \end{minipage}
                              \hspace{1.0in}
                              \begin{minipage}[b]{0.35\linewidth}
                                \begin{center}
                                  \includegraphics[scale=#6]{#5}
                                \end{center}
                                \caption{#7\label{#8}}
                              \end{minipage}
                            \end{center}
                          \end{figure}
                         }

\newcommand{\DoTable}[3]{
                          \begin{table}[ht]
                            \begin{center}
                              #1
                            \end{center}
                            \caption{#2\label{#3}}
                          \end{table}
                         }

\include{figbox}
\bibliographystyle{alpha}

\title{\vspace{-5ex}The Boolean Functions Computed by Random Boolean Formulas\\
OR\\
How to Grow the Right Function}
\author{Alex Brodsky  \and Nicholas Pippenger}
\date{\vspace{-2ex}\small Department of Computer Science,\\
      University of British Columbia,\\
      201-2366 Main Mall, Vancouver, \\
      BC, Canada, V6R 1J9\\
      {\tt \{abrodsky,nicholas\}@cs.ubc.ca}}

\maketitle

\begin{abstract}
Among their many uses, growth processes (probabilistic amplification),
were used for constructing reliable networks from unreliable components,
and deriving complexity bounds of various classes of functions.  Hence,
determining the initial conditions for such processes is an important
and challenging problem.  In this paper we characterize growth processes
by their initial conditions and derive conditions under which results
such as Valiant's\cite{Va84} hold.  First, we completely characterize
growth processes that use linear connectives.  Second, by extending
Savick\'y's~\cite{Sa90} analysis, via ``Restriction Lemmas'', we
characterize growth processes that use monotone connectives, and show
that our technique is applicable to growth processes that use other
connectives as well.  Additionally, we obtain explicit bounds on the
convergence rates of several growth processes, including the growth
process studied by Savick\'y (1990).

\noindent {\bf Keywords:} Computational and structural complexity, growth 
processes, probabilistic amplification
\end{abstract}

\section{Introduction}

The notion of a random Boolean function occurs many times, both
implicitly and explicitly, in the literature of theoretical computer
science.  Not long after Shannon~\cite{Sh38} pointed out the relevance
of Boolean algebra to the design of switching circuits, Riordan and
Shannon~\cite{RiSh42} obtained a lower bound to the complexity (the
size of series-parallel relay circuits, or of formulas with the
connectives ``and'', ``or'' and ``not'') of ``almost all'' Boolean
functions, and this bound can naturally be applied to a ``random''
Boolean function when all $2^{2^n}$ Boolean functions of $n$ arguments
are assumed to occur with equal probability.  Lupanov~\cite{Lu61b}
later showed that Riordan and Shannon's lower bound is matched
asymptotically by an upper bound that applies to all Boolean functions,
so in this situation the average case is asymptotically equivalent to
the worst case.  This asymptotic equivalence of average and worst cases
also holds in many other situations involving circuits or formulas.
There are some complexity measures, however, such as the length of the
shortest disjunctive-normal-form formula, for which the average case
behaves quite differently from the worst case (see Glagolev~\cite{Gl67},
for example), and the complexity of a random Boolean function remains a
challenging open problem.  In these cases, probability distributions
other than the uniform distribution have also been considered; for
example, one may assume that each entry in the truth-table is
independently $1$ with probability $p$ and $0$ with probability $1-p$,
so that the uniform distribution is the special case $p=1/2$ (see
Andreev~\cite{An84}).

Another approach to the study of random Boolean functions is to put a
probability distribution on formulas, and let that induce a probability
distribution on the functions that they compute.  This may be done by
using a ``growth process'' (defined below) to grow random formulas.

Valiant~\cite{Va84} considered such a growth process, and showed that
the resulting probability distribution tends to the distribution
concentrated on a single function:  the threshold function that assumes
the value $1$ if and only at least $n/2$ of its $n$ arguments assume
the value $1$.  This result was used to obtain a non-constructive upper
bound on the minimum possible size of a  formula for computing this
threshold function.  This argument in fact gives the best upper bound
currently known for this and similar threshold functions.

The choice of the initial probability distribution on formulas
dictates the probability distribution on functions.   To facilitate the
design and use of growth process, as in the case above, deriving a
characterization based on the initial conditions is an important problem.

One such result in this framework is due to Savick\'y~\cite{Sa90}.
Savick\'y formulated broad conditions under which the distribution of
the random function computed by a formula of depth $i$ tends to the
uniform distribution on all Boolean functions of $n$ variables as
$i\to\infty$.

Savick\'y~\cite{Sa95} has also shown that in some cases the rate of
approach of the probability of computing a particular function $f$ to
the uniform probability $2^{-2^n}$ gives information about $f$:  it is
fastest for the linear functions, and slowest for the ``bent''
functions (which are furthest, in Hamming distance, from the linear
functions).  For some other models of random formulas, Lefmann and
Savick\'y~\cite{LeSa97} and Savick\'y~\cite{Sa98} have shown that the
logarithm of the probability of computing a particular function is
related to the complexity of that function (as measured by the size of
the smallest formula computing that function).  Finally, we should
mention that Razborov~\cite{Ra88} has used random formulas in yet
another model to show that some large graphs with Ramsey properties
have representations by formulas of exponentially smaller size.  This
result, which has been improved quantitatively by
Savick\'y~\cite{Sa95b}, shows that Ramsey properties are possessed by
graphs that are far from random.

Our goal in this paper is to determine under what circumstances results
like Valiant's and Savick\'y's hold.  We show that for many growth
processes, the probability distribution on the computed function tends
to the uniform distribution on some set of functions (which may range
in size from a single function, as in Valiant's result, to all
functions, as in Savick\'y's).

\section{Definitions}\label{sec:defs}
Let $\cF_n$ denote the family of $n$-adic Boolean functions, let
$\cM_n$ denote the family of $n$-adic monotone Boolean functions, and
let $\cL_n$ denote the family of $n$-adic linear functions.  The set $B_n$
denotes Boolean cube of size $n$.

Let $k$ be a positive integer and $\alpha$ be a $k$-adic Boolean
function, which we call the \myem{connective}.  Let $A_0 =
\{x_1,x_2,...,x_n, \bar{x}_1,...,\bar{x}_n,0,1\}$ be the set comprising
the projection functions, their negations, and the constant functions,
and let $A_i = \{ \alpha(v_1,v_2,...,v_k)\ |\ v_i \in A_{i-1}\}$ be the
set comprising the formulas composed from $A_{i-1}$. A \myem{growth
process} is denoted by a pair $(\mu,\alpha)$, where $\mu$ is a
distribution on $A_0$ and $\alpha$ is a connective; $\mu$ is called the
\myem{initial distribution}.  A growth process gives rise to a
probability distribution $\pi_i$ on $A_i$ for each $i\ge 0$ in the
following way.  We take $\pi_0 = \mu$.  For $i\ge 1$, we take
$\pi_i(f)$ to be the probability that $\alpha(g_1, \ldots, g_k) = f$,
where $g_1, \ldots, g_k$ are independent random functions distributed
according to $\pi_{i-1}$ on $A_{i-1}$.

We shall assume that $\mu$ is a uniform distribution on a subset of
$A_0$.  This subset will always contain the $n$ projections; it may or
may not contain their $n$ negations; and it may contain neither, one,
or both of the two constants.  All of our results could be extended to
more general distributions $\mu$, but these assumptions allow us to
present the most interesting results with a minimum of notation.  They
also cover the results of Valiant~\cite{Va84} and
Savick\'y~\cite{Sa90}.  (Valiant's proof actually uses a non-uniform
distribution, but the same bound can be obtained by a simple
modification using a uniform distribution on the projections.)

The \myem{support} of a probability distribution $\pi$, denoted
$\Supp(\pi)$, is the set $\{f\ |\ \pi(f) > 0\}$.  The \myem{support}
of a growth process is the set of all functions $f\in \cF_n$ for
which $\pi_i(f) > 0$ for some $i > 0$: $\cup_i \Supp(\pi_i)$.

We are particularly interested in cases in which $\pi_i$ tends to
a \myem{limiting distribution} $\pi$ as $i\to\infty$.  (There are
also cases in which $\pi_{2i}$ and $\pi_{2i+1}$ tend to distinct
\myem{alternating limiting distributions}.) When a limiting
distribution exists, we can have $\pi(f) > 0$ only for $f$ in the
support of the growth process.  As Valiant's result indicates,
however, there may be functions in the support for which $\pi_i(f)\to
0$, so that $\pi(f) = 0$.  The \myem{asymptotic support} of a growth
process with a limiting distribution $\pi$ is the set of functions
$f\in \cF_n$ for which $\pi(f) > 0$.

Additionally, we investigate how quickly the distribution $\pi_i$
approaches the limiting distribution as $i$ approaches infinity.
Namely, for some $\epsilon > 0$, the size of $i$ such that
$\max_f|\pi(f) - \pi_i(f)| < \epsilon$.  Almost all growth processes
that we study share the important characteristic:  for any $\epsilon >
0$,\ $\max_f|\pi(f) - \pi_{O(\log(n))}(f)| < \epsilon$.  Note, unless
otherwise stated, the base of the logarithm is assumed to be $2$.

Growth processes in which the limiting distribution is concentrated on
one function are used extensively in probabilistic amplification
methods and can be analyzed by studying the properties of the
corresponding ``characteristic polynomial''.  Let $\{X_1,X_2,...,X_n\}$
be a set of random independent binary variables that are $1$ with
probability $p$ and let each $X_i$ represent the input $x_i$.  The
\myem{characteristic polynomial} of $f$ is defined by $A_f(p) =
\Pr[f(X_1,X_2,...X_n) =1]$ and is given by
 \[A_f(p) = \sum_{i=0}^n \beta_i\Ch{n}{i}p^i(1-p)^{n-i}\]
where $\beta_i$ is the fraction of assignments of weight $i$ for which
$f$ is true.  The characteristic polynomial was used by von
Neumann~\cite{vN56} and by Moore and Shannon~\cite{MoSh56} to study
reliable computation with unreliable components, as well as by
Valiant~\cite{Va84} (see also Boppana~\cite{Bo85,Bo89,DuZw97}).

To analyze growth processes whose limiting distribution is uniform over
a set of functions, we use a Fourier transform technique.  The Fourier
transform $\Delta_i$ of a probability distribution $\pi_i$ is defined by
\begin{equation}\label{eqn:fft}
\Delta_i(f) = \sum_{g \in \cF_n} (-1)^\iprod{f}{g} \pi_i(g)
\end{equation}
where $\pi_i(g)$ is the probability of selecting $g$ from $A_i$.  For
convenience, the inner product $\iprod{f}{g} = \sum_i f_ig_i$ is
defined to be over the integers, rather than over $\Zf_2$.  Unless
otherwise noted, Boolean $n$-adic functions are represented as Boolean
vectors from $B_{2^n}$.  The inverse Fourier transform is defined by
\begin{equation}\label{eqn:inv}
\pi_i(g) = \frac{1}{2^{2^n}}\sum_{f \in \cF_n} (-1)^\iprod{f}{g} \Delta_i(f).
\end{equation}
The Fourier transform was used by Razborov~\cite{Ra88} to derive his
results on Ramsey graphs, as well as by Savick\'y~\cite{Sa90}.

The Fourier transform plays a role in many of our results, but it needs
to be adapted in various ways to suit different cases.  When dealing
with linear functions, for example, we will have to represent the
functions $f$ and $g$ in definition \ref{eqn:fft} not as Savick\'y
does, by their truth-tables, but rather by their coefficients as
multivariate polynomials over $GF(2)$.  In other cases, when
establishing a limiting distribution that is uniform over a proper
subset of $\cF_n$, we shall need to use what we call ``restriction
lemmas'', which assert relationships that hold among the values of the
Fourier transform.

\section{Growing Linear Functions}\label{sec:lin}
A function $f$ is linear if it is of the form $f(x_1, \ldots, x_n) =
c_0 \oplus c_1 x_1 \oplus \cdots \oplus c_n x_n$ for some constants
$c_0, c_1, \ldots, c_n \in GF(2)$.  We may assume without loss of
generality that $\alpha$ depends on all its arguments, so that
$\alpha(y_1, \ldots, y_k) = c \oplus y_1 \oplus\cdots\oplus y_k$, where
$k\ge 2$.  The result of the growth process depends on the support of
of the initial distribution $\mu$, the parity of $k$, and the constant
term $c$.

To prove this we derive a recurrence for the Fourier coefficients of
the respective probability distribution $\pi_i$, from which we derive
the limiting distribution.  Since compositions of linear functions
are themselves linear, we represent the linear functions by their
vector $(c_0, c_1, \ldots, c_n)$ of coefficients, and the following
summations range over $\cL_n$.  Finally, let $w_1$ denote the constant
function $1$ ($w_1 = 100\ldots0$), whereas $\onebf = 11\ldots1$.

\begin{proposition}\label{prop:lin_rec}
Let $\alpha$ be a linear connective as described above and let $w \in \cL_n$.
The Fourier coefficients of the probability distribution $\pi_i$ of the
corresponding growth process are described by the recurrence relation
\[\Delta_{i+1}(w) = (-1)^{c\iprod{w_1}{w}}\Delta_i(w)^k.\]
\end{proposition}
\begin{proof}
\begin{eqnarray*}
\Delta_{i+1}(w) & = & \sum_{f \in \cL_n} \pi_{i+1}(f) (-1)^\iprod{f}{w}
                  =   \sum_{f \in \cL_n}
                      \sum_\Sarg{\gbf \in \cL_n^k}{\alpha(\gbf) = f}
                      \prod_{j=1}^k \pi_i(\gbf_j)(-1)^\iprod{f}{w}\\
                & = & \sum_{\gbf \in \cL_n^k} \prod_{j=1}^k \pi_i(\gbf_j)
                                               (-1)^\iprod{\alpha(\gbf)}{w}
                  =   \sum_{\gbf \in \cL_n^k} \prod_{j=1}^k \pi_i(\gbf_j)
                       (-1)^\iprod{cw_1\oplus\bigoplus_{j=1}^k \gbf_j}{w} \\
                & = & \sum_{\gbf \in \cL_n^k} \prod_{j=1}^k \pi_i(\gbf_j)
            (-1)^{\iprod{cw_1}{w}\oplus\bigoplus_{j=1}^k\iprod{\gbf_j}{w}}
                  =   (-1)^\iprod{cw_1}{w} \sum_{\gbf \in \cL_n^k}
                           \prod_{j=1}^k \pi_i(\gbf_j) (-1)^\iprod{\gbf_j}{w} \\
                & = & (-1)^\iprod{cw_1}{w} \Delta_i(w)^k
\end{eqnarray*}
\end{proof}

Using proposition~\ref{prop:lin_rec}, the following theorems classify the
growth processes on linear connectives.

\begin{theorem}\label{thm:lin}
Let $\alpha(y) = c \oplus y_1 \oplus\cdots\oplus y_k$,\ $k > 1$, be a
linear $k$-adic connective, as defined above, and assume that the support
of $\mu$ does not contain negations of the projections.
\begin{enumerate}
\item If $\{0,1\} \cap \Supp(\mu) \not= \{0,1\}$,\ $k$ is odd and $c = 1$,
      then the growth process has alternating limiting distributions,
      each of which is uniform over one half of the support of the
      growth process (which consists of all linear functions for which
      $\bigoplus_{j=1}^n c_j = 1$).
\item In all other cases, the limiting distribution is uniform over the
      support of the growth process (which depends on $k$, $c$, and the
      presence of constants in the support).
\end{enumerate}
\end{theorem}
\begin{proof}
Two facts are key to this theorem: first, that $|\Delta_i(w)| \leq 1$,
and second, that if $|\Delta_i(w)| < 1$, then $\lim_{i \rightarrow \infty}
\Delta_i(w) = 0$.  Only the nonzero (magnitude 1) coefficients
contribute to limiting distribution (equation~\ref{eqn:inv});
fortunately, these are determined solely by the support of the initial
distribution.  Depending on which constants are part of the support,
there are either one, two, or four magnitude 1 coefficients:
\begin{eqnarray*}
\{0,1\}\cap \Supp(\mu) = \{0,1\} & \Rightarrow & \Delta_0(0) = 1, \\
\{0,1\}\cap \Supp(\mu) = \{0\}   & \Rightarrow & \Delta_0(0)=\Delta_0(w_1)= 1,\\
\{0,1\}\cap \Supp(\mu) = \{1\}   & \Rightarrow & \Delta_0(0) = 1, \
                                                 \Delta_0(\onebf) = -1,\\
\{0,1\}\cap \Supp(\mu) =\emptyset& \Rightarrow & \Delta_0(0) = \Delta_0(w_1)= 1,
                        \ \Delta_0(\onebf)=\Delta_0(w_1\oplus\onebf) = -1.
\end{eqnarray*}

If $k$ is odd and $c = 1$, the recurrence from
Proposition~\ref{prop:lin_rec} implies that $\Delta_{i+1}(w_1) =
-\Delta_i(w_1)$ and $\Delta_{i+1}(\onebf) = -\Delta_i(\onebf)$.  Hence,
if $\Supp(\mu) \cap \{0,1\} \not= \{0,1\}$, the resulting distribution
is alternating.  In the case where one of the constants is missing from
the support, only two coefficients have magnitude 1, and thus, the
alternating distributions are each uniform over half of $\cL_n$.  In
the case where both constants are missing, the alternating
distributions are each uniform over one quarter of $\cL_n$.

If $c = 0$, $k$ is even, or $\{0,1\}\cap \Supp(\mu) = \{0,1\}$, the
limiting distribution exists because the sign of the magnitude 1
coefficients does not alternate.  We can read off the limiting
distribution from the Fourier coefficients.  If both constants are in
the support, then the limiting distribution is uniform over $\cL_n$.
If only one of the constants is present, then the distribution is
uniform over half of $\cL_n$, and if neither is present, then the
distribution will be uniform over a quarter of $\cL_n$.
\end{proof}

If the support of $\mu$ contains negations, then using the same proof
technique yields the following theorem.

\begin{theorem}\label{thm:lin_neg}
Let $\alpha(y) = c \oplus y_1 \oplus\cdots\oplus y_k$ be a linear
$k$-adic connective, as defined above, and assume that the support of
$\mu$ contains negations of the projections.
\begin{enumerate}
\item If $\{0,1\}\cap \Supp(\mu) = \emptyset$ and $k$ is odd then the
      limiting distribution is uniform over all linear functions of odd
      number of variables.
\item If $\{0,1\}\cap \Supp(\mu) = \emptyset$ and $k$ is even then the
      limiting distribution is uniform over all linear functions of
      even number of variables.
\item Otherwise, the limiting distribution is uniform over all of $\cL_n$.
\end{enumerate}
\end{theorem}
\begin{proof}
If $\{0,1\}\cap \Supp(\mu) \not= \emptyset$, then there is only one
coefficient of magnitude $1$, $\Delta_0(\zerobf) = 1$, implying the
last case.

Otherwise, there is one other magnitude $1$ coefficient,
$\Delta_0(\onebf \oplus w_1) = -1$.  If $k$ is odd, then
$\Delta_{i+1}(\onebf \oplus w_1) = \Delta_i(\onebf \oplus w_1)^k = -1$,
implying the first case of the theorem.  If $k$ is even, then
$\Delta_{i+1}(\onebf \oplus w_1) = \Delta_i(\onebf \oplus w_1)^k = 1$,
implying the second case.
\end{proof}

Note, that if negations are present, no alternating distribution can occur.
To bound the convergence of $\pi_i$ to $\pi$ we use the inverse Fourier
transform.

\begin{theorem}
Let $\alpha$ be a $k$-adic linear connective, $k > 1$, of a linear
process on $n$ variables that has a limiting distribution $\pi$.  If 
$i > \frac{2\log(n)}{\log(k)}$, then for any linear function $f$, 
$|\pi(f) - \pi_i(f)| < 2^{-n}$.
\end{theorem}
\begin{proof}
Let $D = \{w\ :\ |\Delta_0(w)| < 1\}$, then $\pi_i(f)$ may be written as:
\begin{eqnarray*}
\pi_i(f) & = & 2^{-n-1}\sum_{w \in B_{n+1}}(-1)^\iprod{w}{f}\Delta_i(w) \\
         & = & 2^{-n-1}\sum_{w \not\in D}(-1)^\iprod{w}{f}\Delta_i(w)
             + 2^{-n-1}\sum_{w \in D}(-1)^\iprod{w}{f}\Delta_i(w)\\
         & = & \pi(f)
             + 2^{-n-1}\sum_{w \in D}(-1)^\iprod{w}{f}\Delta_i(w).
\end{eqnarray*}
Thus, for any linear function $f$,
\[|\pi(f) - \pi_i(f)| = |2^{-n-1}\sum_{w \in D}(-1)^\iprod{w}{f}\Delta_i(f)|
   \leq   \max_{w \in D}|\Delta_0(w)|^{k^i} \leq (1-n^{-1})^{k^i}.\]
Solving inequality $(1-n^{-1})^{k^i} < 2^{-n}$, in terms of $i$, yields:
$i > \frac{2\log(n)}{\log(k)}$.
\end{proof}

\section{Growing Self-Dual Functions}\label{sec:self}
Savick\'y~\cite{Sa90} showed that if the connective is balanced (that
is, if it assumes the value $1$ for just one-half of the combinations
of argument values) and non-linear, and the support of $\mu$ is all of
$A_0$, then the limiting distribution will be uniform over all of
$\cF_n$.  If we remove the constants from the support of $\mu$ and
assume the connective $\alpha$ is self-dual (that is, satisfies
$\alpha(y_1, \ldots y_k) = \overline{\alpha(\bar{y}_1, \ldots,
\bar{y}_k)}$), then the support of the growth process is the set of all
self-dual functions.  In this case the limiting distribution of the
growth process is uniform over this support.

\begin{theorem}\label{thm:self_dual}
If the connective is non-linear and self-dual, and the support of $\mu$
comprises the projections and their negations, then the limiting
distribution will be uniform over the family of self-dual $n$-adic
functions.
\end{theorem}
\begin{proof}
Observe that there is a bijection between the set of all functions on $n$
variables and the set of self-dual functions on $n+1$ variables, for
example, the map
\[f(x_1,x_2,\ldots,x_n) \mapsto f(x_1,x_2,\ldots,x_n)x_{n+1}
  \vee \overline{f(\bar{x}_0,\bar{x}_1,\ldots,\bar{x}_n)}\bar{x}_{n+1}.\]
The result follows.
\end{proof}

\section{Growing Monotone Functions}\label{sec:mono}
We now focus on growth processes that use monotone connectives.  For
the rest of this section we assume that $\alpha$ is monotone and the
support of $\mu$ contains only monotone functions from $A_0$ (that is,
projections and possibly constants).  We first investigate unbalanced
connectives.

\subsection{Using Unbalanced Connectives}
Growth processes that use unbalanced monotone connectives concentrate
probability on a threshold function; the type of threshold function
depends on the connective and the support.  A threshold function
$T_k(x_1, \ldots, x_n)$ assumes the value $1$ if and only if at least
$k$ of its $n$ arguments assume the value $1$.  We consider constant
functions $T_{n+1} = 0$ and $T_0 = 1$ to be special cases of threshold
functions.  There are two cases to consider: first, when the
characteristic polynomial of $\alpha$, $A_\alpha(p)$, has no fixed-point
on the open interval $(0,1)$, and second, when $A_\alpha(p)$ has a
fixed-point on $(0,1)$.

\begin{proposition}\label{prop:no_fp}
If $\alpha$ is a monotone connective whose characteristic polynomial,
$A(p)$, has no fixed-point on the interval $(0,1)$, then the limiting
distribution will be concentrated on a threshold function.
\end{proposition}
\begin{proof}
Since $A(p)$ has no fixed-point on $(0,1)$, either $A_\alpha(p) < p$
throughout $(0,1)$, or $A_\alpha(p) > p$ throughout $(0,1)$.  If
$A_\alpha(p) > p$ throughout $(0,1)$, then by the standard
amplification argument, the limiting distribution is concentrated on
$T_1$ (disjunction of all variables) or, $T_0$  if $1$ is in the
support of $\mu$.  Similarly, if $A_\alpha(p) < p$ throughout $(0,1)$,
then the limiting distribution is concentrated on $T_n$ (conjunction of
all variables) or $T_{n+1}$ if $0$ is in the support of $\mu$.
\end{proof}

Furthermore, all connectives whose characteristic polynomials have no
fixed-point on $(0,1)$, are either of the form $\alpha(x) = x_i \vee
\alphap(x)$ (when $A_\alpha(p) > p$) or $\alpha(x) = x_i \wedge
\alphap(x)$ (when $A_\alpha(p) < p$).  If $\alpha(x) \not= x_i \vee
\alphap(x)$, then $A_\alpha(p) = O(p^2)$ which implies that there
exists a positive constant $\epsilon_0$ such that for all $0 < \epsilon
< \epsilon_0$, $A_\alpha(\epsilon) < \epsilon$.  Similarly, if
$\alpha(x) \not= x_i \wedge \alphap(x)$, then by duality, $1 -
A_\alpha(1-p) = O(p^2)$, which means that $A_\alpha(1-\epsilon) >
1-\epsilon$ for all $0 < \epsilon < \epsilon_1$ for some $\epsilon_1 >
0$.  Since $A_\alpha(p)$ is continuous, there must exist a
fixed-point in $(0,1)$, which is a contradiction.

In the second case, where $A_\alpha(p)$ has a fixed-point in $(0,1)$,
Moore and Shannon~\cite{MoSh56} have shown that this fixed-point is
unique.  Not surprisingly, the limiting distribution depends on the
fixed-point.  Thus, we first derive two facts about the fixed-point of
the characteristic polynomial, to deal with the second case.

\begin{lemma}\label{lem:bal}
The characteristic polynomial $A(p)$ has a fixed-point of $\Half$ if
and only if the connective $\alpha$ is balanced.
\end{lemma}
\begin{proof}
By definition $\sum_{i=0}^n \beta_i\Ch{n}{i}$ is the number of
assignments for which $\alpha$ is true.  If $A_\alpha(\Half) = \Half$, then
$A_\alpha(\Half) = \sum_{i=0}^n \beta_i\Ch{n}{i} (\Half)^i(\Half)^{n-i} =
\frac{1}{2^n} \sum_{i=0}^n \beta_i\Ch{n}{i} = \Half$.  Hence,
$\sum_{i=0}^n \beta_i \Ch{n}{i} = 2^{n-1}$ which means that $\alpha$ is
balanced.  Conversely, if $\alpha$ is balanced, then $A_\alpha(\Half) =
\Half$.
\end{proof}

\begin{lemma}\label{lem:equiv}
If $\alpha$ is a monotone, non-projection connective, then any fixed-point
of $A_\alpha(p)$ on $(0,1)$ is either irrational or $\Half$.
\end{lemma}
\begin{proof}
By contradiction; without loss of generality assume that the fixed-point
$p_0 = \frac{r}{s} < \frac{1}{2}$ and $\gcd{(r,s)} = 1$.  Hence,
\[A_\alpha\left(\frac{r}{s}\right) = \sum_{j=0}^k \beta_j\Ch{k}{j}
   \left(\frac{r}{s}\right)^j\left(\frac{s - r}{s}\right)^{k-j} =
\frac{r}{s}.\]
Multiplying both sides by $s^k$, noting that $\beta_k = \beta_{k-1} =
1$, and evaluating the result modulo $(s-r)^2$ yields
\begin{eqnarray*}\label{eq:fp_half}
rs^{k-1} & \equiv & \sum_{j=0}^k \beta_j\Ch{k}{j} r^j(s - r)^{k-j}
          \equiv  r^k + kr^{k-1}(s-r) +
                (s-r)^2\sum_{j=0}^{k-2}\beta_j\Ch{k}{j}r^j(s-r)^{k-j-2}\\
         & \equiv & r^k + kr^{k-1}(s-r) \bmod{(s-r)^2}.
\end{eqnarray*}
Evaluating the left side modulo $(s-r)^2$ yields
\begin{eqnarray*}
rs^{k-1} & \equiv & r(r + (s-r))^{k-1}
           \equiv  r\sum_{i=0}^{k-1} \Ch{k-1}{i}r^i(s-r)^{k-1-i} \\
         & \equiv & rr^{k-1} + r(k-1)r^{k-2}(s-r) +
                    r(s-r)^2\sum_{j=0}^{k-3}\Ch{k-1}{j}r^j(s-r)^{k-3-j} \\
         & \equiv & r^k + (k-1)r^{k-1}(s-r) \bmod{(s-r)^2}.
\end{eqnarray*}
Therefore,
\[r^{k-1}(s-r) \equiv 0 \bmod{(s-r)^2}.\]
Since $\gcd{(r,s)} = \gcd{(r,(s-r)^2)} = 1$, $r^{k-1} \not\equiv 0
\bmod{(s-r)^2}$; this is a contradiction.
\end{proof}

\begin{theorem}\label{thm:mono_unbal}
Let $\alpha$ be a monotone unbalanced connective whose characteristic
polynomial has a fixed-point $t \in (0,1)$, and let the support of
$\mu$ contain only the projections.  The limiting distribution of
the growth process is concentrated on the threshold function $T_{\lceil
tn\rceil}$.
\end{theorem}
\begin{proof}
Since $\alpha$ is unbalanced and has a fixed-point on $(0,1)$, by
Lemma~\ref{lem:bal}, the fixed-point is not $\Half$.  Hence, by
Lemma~\ref{lem:equiv}, the fixed-point is irrational.  Since the
fraction of variables set to true in any assignment is by definition
rational, the fraction will always be strictly greater or strictly less
than the fixed-point $t$.  Hence, by the standard amplification argument,
the limiting distribution will be concentrated on the threshold function
$T_{\lceil tn\rceil}$.
\end{proof}

Theorem \ref{thm:mono_unbal} can easily be modified to cover the cases
in which one or both constants are in the support of $\mu$.   Combining
proposition~\ref{prop:no_fp} and theorem~\ref{thm:mono_unbal} proves
the initial claim.

\begin{theorem}
If $\alpha$ is a monotone unbalanced connective and the support of
$\mu$ does not contain the negations of projections, then the limiting
distribution will be concentrated on a threshold function.
\end{theorem}

\subsubsection{Convergence Bounds}
Except in one case, all these growth processes converge very quickly to
their limiting distribution: in $O(\log(n))$ iterations.  In the
exceptional case the convergence requires $O(n^k)$ iterations where $k$
is the arity of the connective $\alpha$; we provide specific criteria
that determine whether a process will converge quickly or not.  There
are two main cases: either $A_\alpha(p)$ has a fixed-point, or not.  We
first derive bounds for the latter case, and then for the former.
Unless explicitly stated, we assume that constants are not in
$\Supp(\mu)$, however, the following analysis changes little if
constants are in $\Supp(\mu)$.

In the first case, either $A_\alpha(p) > p$ for $p \in (0,1)$, and
$A_\alpha(p) < p$, for $p \in (0,1)$.  Since, the two cases are
symmetric, the same bounds apply to both.  Hence, without loss of
generality assume that $A_\alpha(p) < p$ on the interval $(0,1)$.

\begin{lemma}\label{LEM:FAST-NFP}\label{lem:fast-nfp}
If $\alpha$ is a monotone connective such that $A_\alpha(p) < p$ on the
interval $(0,1)$ and, $A_\alpha(p)$ has degree $k > 2$ and $\beta_{k-1}
\leq \frac{k-2}{k}$, then for all positive $\epsilon < \epsilon_k =
\frac{1}{k2^{k+1}}$,
\[\frac{35}{24} < \Ap_\alpha(1-\epsilon)\]
\end{lemma}
\begin{proof}
See Appendix.
\end{proof}

\begin{lemma}\label{LEM:SLOW-NFP}\label{lem:slow-nfp}
If $\alpha$ is a monotone connective such that $A_\alpha(p) < p$ on the
interval $(0,1)$ and, $A_\alpha(p)$ has degree $k > 2$ and $\beta_{k-1}
= \frac{k-1}{k}$, then, for all positive $\epsilon < k^{-1}$,
\[1+\epsilon^k < \Ap_\alpha(1-\epsilon)
            \leq (1-\epsilon)^{k-2}(k(k-2)\epsilon+1)\]
\end{lemma}
\begin{proof}
See Appendix.
\end{proof}

\begin{lemma}\label{LEM:TRIV}\label{lem:triv}
If $\alpha$ is a monotone connective that is not of the form $\alpha(x)
= x_i \vee \alphap(x)$, then on the interval $(0,1)$, $A_\alpha(p) <
(\Ch{k}{2}+1)p^2$.
\end{lemma}
\begin{proof}
See Appendix.
\end{proof}

\begin{theorem}
Let $\alpha$ be a $k$-adic monotone connective such that $A_\alpha(p) <
p$ on the interval $(0,1)$, $k > 2$ and $\beta_{k-2} \leq \frac{k-2}{k}$.
There exists a constant $c_\alpha$, such that for all $n > 0$, if $i \geq
3\log(n) + c_\alpha$, then for all $f$, $|\pi_i(f) - \pi(f)| < 2^{-n}$.
\end{theorem}
\begin{proof}
Let $\ft_i$ be a random variable with the distribution $\pi_i$.  Using
an argument similar to Valiant's~\cite{Va84}, we claim that if $i \geq
3\log(n) + c_\alpha$, then for $|x| = n$, $P[\ft_i(x) = 0] = 0$, and
for all $x$ such that $|x| < n$, $P[\ft_i(x) = 1] < 2^{-2n}$.  The
former follows from the monotonicity of $\alpha$; regardless of the
number of iterations, a false negative will never occur.

In the latter case, assuming that all variables are independent, if
$|x| < n$, then $P[\ft_0(x) = 1] = |x|/n \leq 1 - n^{-1}$.  For $i > 0$,
$P[\ft_i(x) = 1] =  A_\alpha^i(p)$, where $A_\alpha^i$ denotes the $i$th
composition of $A_\alpha$ with itself.  Expanding $A_\alpha(p)$ around $1$,
\[A_\alpha(p) = A_\alpha(1) + \Ap_\alpha(1)(p-1) + O((p-1)^2),\]
yields:
\[A_\alpha(1 - \epsilon) = 1 - \epsilon\Ap_\alpha(1) + O(\epsilon^2).\]
From Lemma~\ref{lem:fast-nfp}, let $\gamma = 35/24$ and let $\epsilon_k =
\frac{1}{k2^{k+1}}$.  There exists an $\epsilon_0 < \epsilon_k$ such that
for all $\epsilon < \epsilon_0$
\[A_\alpha(1 - \epsilon) < 1 - \epsilon\gamma.\]
Since $P[\ft_0(x) = 1] \leq 1 - n^{-1}$, for $i \geq 2\log(n) +
2\log(\epsilon_0) > (\log(n) + \log(\epsilon_0))/\log(35/24)$,
\[A_\alpha^i(1 - \epsilon) < 1 - \epsilon\gamma^i < 1 - \epsilon_0.\]

An additional constant number of iterations, say $d_\alpha$, yields
\[A_\alpha^{d_\alpha}(1 - \epsilon_0) < c.\]
By Lemma~\ref{lem:triv}, $A_\alpha(p) < k^2p^2$, thus we fix $c <
\frac{1}{2k^2}$ and let $j = \log(n) + 1$. Hence,
\[A_\alpha^j(c) < (k^2c)^{2^j} < 2^{-2^j} = 2^{-2n}.\]

Therefore, for $i \geq 3\log(n) + 2\log(\epsilon_0) + d_\alpha + 1$ and
all $x$ such that $|x| < n$,\ $P[\ft_i(x)=1]<2^{-2n}$, implying that
$|\pi_i(f) - \pi(f)| < 2^{-n}$.
\end{proof}

Unfortunately, if $\beta = \frac{k-1}{k}$, convergence takes time
polynomial in $n$.  If $|x| = n-1$ then $P[\ft_0(x) = 1] = 1 - n^{-1}$.
Furthermore, by Lemma~\ref{lem:slow-nfp}, for sufficiently large $n$,
$\Ap_\alpha(1-n^{-1}) < (1-n^{-1})^{k-2}(k(k-2)n^{-1} + 1)$.  Since
$\gamma < \Ap_\alpha(1-n^{-1})$, therefore
\[\log(\gamma) < (k-2)\log(1-n^{-1}) + \log(k(k-2)n^{-1} + 1),\]
implying that
\[\log(\gamma)^{-1} > \left((k-2)\log(1-n^{-1})+\log(k(k-2)n^{-1}+1)\right)^{-1}
                    > \frac{n}{k^2-3k+2} + O(1).\]
Thus, if $A_\alpha^i(1-n^{-1}) < 1-\epsilon_0$, then for sufficiently
large $n$, $A_\alpha^i(1-n^{-1}) < \epsilon_0$ implies that $i >
(k-1)(k-2)2n(\log(n) + \log(\epsilon_0))$.  In fact, this is the best
case.  If $\alpha(x) = \vee_{i=2}^k (x_1\wedge x_i)$, then $A_\alpha(p)
= p - p(1-p)^{k-1}$.  By Lemma~\ref{lem:slow-nfp}, $\gamma < 1 +
n^{2-k}(k-1-kn^{-1})$, implying that $\log(\gamma) < \log(1 +
n^{2-k}(k-1-kn^{-1}))$, and
\[\log(\gamma)^{-1} > \left(\log( 1 + n^{2-k}(k-1-kn^{-1}))\right)^{-1}
                    > \frac{n^{k-2}}{k-1}.\]
Thus, if $A_\alpha^i(1-n^{-1}) < 1-\epsilon_0$, then for sufficiently
large $n$, $i > \frac{n^{k-2}}{k-1}(\log(n) + \log(\epsilon_0))$.
Consequently, connectives whose characteristic polynomial has no
fixed-point can be classified as either quickly converging or slowly
converging, with the value of the second last coefficient,
$\beta_{k-1}$, determining rate of convergence!

When the characteristic polynomial $A_\alpha(p)$ does have a
fixed-point on the interval $(0,1)$, a similar analysis is used.

\begin{lemma}\label{LEM:MINSLOPE}\label{lem:minslope}
Let $A_\alpha(p)$ be the characteristic polynomial of any $k$-adic
monotone connective. If $A_\alpha(p)$ has a fixed-point $s \in (0,1)$,
then $\Ap_\alpha(s) \geq 1 + \frac{k-2}{2^{k-2}}$.
\end{lemma}
\begin{proof}
See Appendix.
\end{proof}

\begin{theorem}\label{thm:bound-fp}
Let $\alpha$ be a $k$-adic monotone connective such that $A_\alpha(s) =
s \in (0,1)$.  There exists a constant $c_\alpha$, such that for all $n
> 0$, if $i \geq k2^k\log(n) + c_\alpha$, then for all functions
$f$, $|\pi_i(f)-\pi(f)|< 2^{-n}$.
\end{theorem}
\begin{proof}
Let $\ft_i$ be a random variable with the distribution $\pi_i$.  Using
an argument similar to Valiant's~\cite{Va84}, we claim that if $i \geq
k2^k\log(n) + c_\alpha$ then for all $x$ such that $|x| < sn$,
$P[\ft_i(x) = 1] < 2^{-2n}$, and for all $x$ such that $|x| > sn$,
$P[\ft_i(x) = 0] < 2^{-2n}$.  We first argue the former.

Assuming that all variables are independent, if $|x| < sn$, then $P[\ft_0(x)
= 1] \leq s - n^{-1}\epsilon_\alpha(n)$, where $\epsilon_\alpha(n) =
\min_{j\in\Zf}|s - \frac{j}{n}| = |s - \frac{j_0}{n}|$.  Since $s$ is
an algebraic of degree $k$, by Liouville's Approximation
Theorem~\cite{Ap97}
\[\epsilon_\alpha(n) = \left|s - \frac{j_0}{n}\right| > \frac{e_\alpha}{n^k},\]
where the constant $e_\alpha$ depends only on the connective.

For $i > 0$, $P[\ft_i(x) = 1] = A_\alpha^i(p)$, where $A_\alpha^i$
denotes the $i$th composition of $A_\alpha$ with itself.  Expanding
$A_\alpha(p)$ around $s$,
\[A_\alpha(p) = A_\alpha(s) + \Ap_\alpha(s)(p-s) + O((p-s)^2),\]
yields:
\[A_\alpha(s - \epsilon) = s - \epsilon\Ap_\alpha(s) + O(\epsilon^2).\]
By Lemma~\ref{lem:minslope}, fix $\gamma = 1+2^{-k+1}$; there
exists an $\epsilon_0$ such that for all $\epsilon < \epsilon_0$,
$A_\alpha(s - \epsilon) < s - \epsilon\gamma$.  Since $P[\ft_0(x) = 1]
\leq s - n^{-1}\epsilon_\alpha(n)$, if
\[i \geq \log(n\ \epsilon_\alpha(n)^{-1}\epsilon_0)/\log(\gamma) \geq
         \log(n^{k+1}e_\alpha^{-1}\epsilon_0)/\log(\gamma),\]
then
\[A_\alpha^i(s - \epsilon) < s - \epsilon\gamma^i < s - \epsilon_0.\]

An additional constant number of iterations, say $d_\alpha$, yields
\[A_\alpha^{d_\alpha}(s - \epsilon_0) < c;\]
By Lemma~\ref{lem:triv}, $A_\alpha(p) < k^2p^2$, thus, we fix $c <
\frac{1}{2k^2}$ and let $j = \log(n) + 1$. Hence,
\[A_\alpha^j(c) < (k^2c)^{2^j} < 2^{-2^j} = 2^{-2n}.\]
Therefore, if
\[i \geq k2^k\log(n) + \frac{\log(e_\alpha^{-1}\epsilon_0)}{\log(\gamma)}
                            + d_\alpha + 1,\]
for all $x$ such that $|x| < sn$, $P[\ft_i(x) = 1] < 2^{-2n}$.

By the same argument, if $|x| > sn$, $P[\ft_i(x) = 0] < 2^{-2n}$. Since
$|x| > sn$, for $P[\ft_0(x) = 0] \leq 1 - s - n^{-1}\epsilon_\alpha(n)$,
$P[\ft_1(x) = 0] = \bar{A}_\alpha(p) = 1 - A_\alpha(1 - p)$, and
$P[\ft_j(x) = 0] = \bar{A}_\alpha^j(p)$, $j > 0$.  Just as in the
preceding case, the composition of $\bar{A}_\alpha$ with itself first
yields a first order divergence from $1 - s$, followed by a second order
convergence towards zero.  Therefore, $|\pi_i(f) - \pi(f)| < 2^{-n}$.
\end{proof}

To reduce the constant in front of the log term, one solution is to use
a non-uniform initial distribution, as was done by Valiant~\cite{Va84}.

\subsection{Using Balanced Connectives}
In this subsection, it will be convenient to start by assuming that the
support of $\mu$ contains both constants, as well as the projections,
and to deal later with the cases in which one or both constants are
missing from the support of $\mu$.  If the connective is balanced, then
by Lemma~\ref{lem:bal}, its characteristic polynomial has a fixed-point
of $\Half$.  If the number $n$ of variables is odd, then the fraction
of inputs that are true for any assignment is bounded away from
$\Half$, that is, for any $j \in \{1,2,..,n+1\}, |\Half -
\frac{j}{n+2}| \geq \frac{1}{2n+4}$.  Hence, by the standard
amplification argument, the limiting distribution will be concentrated
on the $n$-adic majority function $T_{\lceil n/2\rceil}$.  In fact, the
convergence to the majority function is logarithmic in $n$; by
Theorem~\ref{thm:bound-fp}, if $i \geq k2^k\log(n) + O(1)$, the
$|\pi_i(f) - \pi(f)| < 2^{-n}$.  When the number of variables is even,
however, something completely different happens.

The family of slice functions, denoted $\cS_{m,n}$ and defined by
Berkowitz~\cite{Ber82}, are monotone $n$-adic functions that assume the
value $1$ for all assignments of weight greater than $m$, assume the
value $0$ for all assignments of weight less than $m$, and may take on
either value for assignments of weight $m$.  Unlike other growth
processes where the distribution is either concentrated on a single
function or is uniform on the support of the growth process, the growth
processes we are about to deal with have a limiting distribution that
is uniform on  $\cS_{n/2,n}$.  This set includes a large number,
$2^{n\choose n/2}$, of functions; but according to a result of
Korshunov~\cite{Ko80}, includes only a tiny fraction, less than $\exp
-({n\choose n/2+1} 2^{-n/2})$, of the support of the growth process,
which is the set $\cM_n$ of all monotone functions.

Define the $n$-adic functions
\[\chih(x) = \left\{\begin{array}{lr}
                    1, & |x| = \frac{n}{2} \\
                    0, & \mathit{otherwise}
                    \end{array}\right.
\mathrm{\ \ \ \ and\ \ \ \ }
\upsh(x) = \left\{\begin{array}{lr}
                    1, & |x| > \frac{n}{2} \\
                    0, & \mathit{otherwise}
                    \end{array}\right..\]

\begin{claim}\label{thm:Fourier_slice}
The Fourier coefficients of the probability distribution $\pi$ that is
uniform on the slice functions in $\cS_{n/2,n}$ are given by
\[\Delta(f) = \left\{\begin{array}{lr}
                     0,    & \iprod{f}{\chih} \not=0 \\
                     (-1)^\iprod{f}{\upsh}, & \iprod{f}{\chih} = 0
                     \end{array}\right.\]
\end{claim}
\begin{proof}
Let $c = {|\Shalf|}^{-1} = 2^{-{n\choose n/2}}$.  If $\iprod{f}{\chih} = 0$,
then
\[\Delta(f)   =   \sum_{g \in \cF_n} (-1)^\iprod{f}{g} \pi(g)
              =   c\sum_{g \in \Shalf} (-1)^\iprod{f}{g}
              =   c\sum_{g \in \Shalf} (-1)^\iprod{f}{\upsh}
              =   (-1)^\iprod{f}{\upsh}.\]
Otherwise let $w$ be a singleton such that $w \leq f \wedge \chih$ and
let $W = \{g \in \Shalf\ |\ g \geq w\}$.  Then
\[\Delta(f) = c\sum_{g \in \Shalf} (-1)^\iprod{f}{g}
            = c\sum_{g \in W} (-1)^\iprod{f}{g}+(-1)^\iprod{f}{g \oplus w}
            = 0.\]
\end{proof}

We shall need to combine amplification with Fourier methods to obtain
our result in this case.  The following ``Restriction Lemma'' is the
key to doing this.

\begin{claim}
Let $\alpha$ be a balanced monotone connective.  Then if $f(x) = 1$ for
some $x$ with $|x| < n/2$, or if $f(x) = 0$ for some $x$ with $|x| > n/2$,
then $\lim_{i\rightarrow\infty} \pi_i(f) = 0$.
\end{claim}
\begin{proof}
This follows from Theorem~\ref{thm:bound-fp}.
\end{proof}

\begin{lemma}[The Restriction Lemma]\label{lem:restrict}\ \\
If $\alpha$ is a balanced monotone connective, then for all $w \in \cF_n$,
\[\lim_{i\rightarrow\infty} \Delta_i(w) =
   (-1)^\iprod{\upsh}{w} \lim_{i\rightarrow\infty} \Delta_i(w \wedge \chih).\]
\end{lemma}
\begin{proof}
\newcommand{\wcxh}{{w\cap\chih}}
We begin with the definition
\[\Delta_i(w) = \sum_{v \in B_n} (-1)^\iprod{v}{w} \pi_i(v),\]
then rewrite the equation as
\[\Delta_i(w) = \sum_{v \in B_n} (-1)^\iprod{v}{w}\pi_i(v)
              = \sum_{t \leq \chih}\sum_{u \leq \notchih}
                   (-1)^\iprod{t\vee u}{w} \pi_i(t \vee u),\]
and consider the restriction of $w$ to the slice $\frac{n}{2}$, that is,
$w \wedge \chih$.
Since $\lim_{i\rightarrow\infty} \pi_i(t \vee u) = 0$ if $u \not= \upsh$,
$\lim_{i\rightarrow\infty} \Delta_i(w\wedge \chih)$ can be rewritten as
\begin{eqnarray*}
\lim_{i\rightarrow\infty} \Delta_i(w\wedge \chih)
 &=& \lim_{i\rightarrow\infty} \sum_{v \in B_n}
             (-1)^\iprod{v}{w\wedge \chih}\pi_i(v) \\
 &=& \lim_{i\rightarrow\infty} \sum_{t\leq\chih}\sum_{u \leq \notchih}
                   (-1)^\iprod{t\vee u}{w\wedge \chih} \pi_i(t \vee u) \\
 &=& \sum_{t\leq\chih}\sum_{u \leq \notchih}(-1)^\iprod{t\vee u}{w\wedge \chih}
                                   \lim_{i\rightarrow\infty}\pi_i(t \vee u) \\
 &=& \sum_{t \leq \chih} (-1)^\iprod{t \vee \upsh}{w\wedge \chih}
                                 \lim_{i\rightarrow\infty}\pi_i(t \vee \upsh) \\
 &=& \sum_{t \leq \chih} (-1)^\iprod{t}{w\wedge \chih}
                                 (-1)^\iprod{\upsh}{w\wedge \chih}
                                 \lim_{i\rightarrow\infty}\pi_i(t \vee \upsh)\\
 &=& \sum_{t \leq \chih} (-1)^\iprod{t}{w\wedge \chih}
                                 \lim_{i\rightarrow\infty}\pi_i(t \vee \upsh) \\
 &=& \sum_{t \leq \chih} (-1)^\iprod{t}{w}
                                   \lim_{i\rightarrow\infty}\pi_i(t \vee \upsh).
\end{eqnarray*}
This, in conjunction with
\begin{eqnarray*}
\lim_{i\rightarrow\infty} \Delta_i(w)
 &=& \lim_{i\rightarrow\infty} \sum_{t\leq\chih}\sum_{u \leq \notchih}
                   (-1)^\iprod{t\vee u}{w} \pi_i(t \vee u)
  =  \sum_{t\leq\chih}\sum_{u \leq \notchih} (-1)^\iprod{t\vee u}{w}
                                 \lim_{i\rightarrow\infty} \pi_i(t \vee u) \\
 &=& \sum_{t \leq \chih} (-1)^\iprod{t \vee \upsh}{w}
                                 \lim_{i\rightarrow\infty} \pi_i(t \vee \upsh)
  =  \sum_{t \leq \chih} (-1)^\iprod{t}{w}(-1)^\iprod{\upsh}{w}
                                 \lim_{i\rightarrow\infty}\pi_i(t \vee \upsh) \\
 &=& (-1)^\iprod{\upsh}{w} \sum_{t \leq \chih} (-1)^\iprod{t}{w}
                                 \lim_{i\rightarrow\infty} \pi_i(t \vee \upsh)
  =  (-1)^\iprod{\upsh}{w} \lim_{i\rightarrow\infty} \Delta_i(w\wedge \chih)
\end{eqnarray*}
yields the identity.
\end{proof}

Hence, all we need to show is that $\lim_{i\rightarrow\infty}\Delta_i(w)
= 0$ for $w$ such that $0 < w \leq \chih$. To do this we use
Savick\'y's~\cite{Sa90} argument, which uses induction on the weight of $w$
together with the recurrence
\[\Delta_{i+1}(w) = \sum_{j=1}^k a_j(w)\Delta_i(w)^j + y_i(w)\]
where
\begin{eqnarray*}
a_j(w) & = & \sum_\Sarg{t \in B_k}{|t| = j} S_\alpha(t)^{|w|}, \\
y_i(w) & = & \sum_\Sarg{\vbf \in \cF_n^k}{0 < \vbf_j < w}
                \prod_\Sarg{a \in B_n}{w(a) = 1} S_\alpha(\vbf(a))
                \prod_{j=1}^k \Delta_i(\vbf_j), \\ % \mathrm{and} \\
S_\alpha(t)
       & = & \frac{1}{2^k}\sum_{r \in B_k} (-1)^\iprod{r}{t}(-1)^{\alpha(r)}.
\end{eqnarray*}
Since the connective is not linear, this recurrence is much more
complicated than the one in proposition~\ref{prop:lin_rec}.  The
result is the following proposition.

\begin{proposition}\label{thm:result}
Let $\alpha$ be a monotone balanced non-projection connective, $n$ be even
and the support of $\mu$ comprise the projection functions and constants.
If $0 < w \leq \chih$, then $\lim_{i \rightarrow \infty} \Delta_i(w) = 0$.
\end{proposition}
\begin{proof}
Let $w \leq \chih$ and recall equation~\ref{eqn:fft}:
\[\Delta_0(w) = \sum_{f \in F_n} \pi_0(f)(-1)^\iprod{f}{w}
              = \frac{1}{n +2}\sum_{f \in A_0}(-1)^\iprod{f}{w}.\]
To prove this proposition we need only show that $\Delta_0(w) = 0$ if
$|w| = 1$, and that $|\Delta_0(w)| < 1$ if $|w| = 2$.  In the first
case, since $w \leq \chih$, $w$ is true on a single assignment of
weight $n/2$.  Hence, $\iprod{w}{x_i} = 1$ for exactly half the
projections, where $x_i$ is the $i$th projection function.   Hence, the
projections cancel each other out.  Similarly, the two constants
annihilate one another.  Hence, $\Delta_0(w) = 0$ if $|w| = 1$.

The latter case, $|w| = 2$, is only slightly harder.  Since the
constant $0$ is part of the support, there will at least one positive
contribution, $-1^\iprod{\zerobf}{w}\pi_0(\zerobf) = \frac{1}{n+2}$.
Hence, in order for $|\Delta_0(w)| = 1$ all other contributions must
also be positive, specifically, $\iprod{w}{x_i} = 0$ for all $x_i$; by
the pigeonhole principle this is not possible.  Hence, $|\Delta_0(w)| <
1$ if $|w| = 2$.  Substituting these base cases into Theorem 5.3
of~\cite{Sa90} yields the result.  
\end{proof}

This proposition, together with Claim~\ref{thm:Fourier_slice}, yields the
one of our main results.

\begin{theorem}
Let $\alpha$ be a monotone balanced non-projection connective, $n$ be even
and the support of $\mu$ comprise the projection functions and constants.
Then the limiting distribution is uniform on the functions in $\cS_{n/2,n}$.
\end{theorem}

\subsubsection{Convergence Bounds}
To bound the convergence within the slice we use a theorem of
Savick\'y~\cite{Sa95}; the conditions of the theorem are verified in
Theorem~\ref{thm:result}.

\begin{theorem}[(Savick\'y, 4.8 in \cite{Sa95})] \label{thm:t48}
If $\alpha$ is balanced and nonlinear, $\Delta_0(w) = 0$ for every $w$
such that $|w| = 1$, $\Delta_0(w) < 1$ for every $w$ such that $|w| =
2$, and there exists a $w$ such that $|w| = 2$ and $\Delta_0(w) > 0$,
then
\[\max_{f \in \cF_n}{|\pi_i(f) - \pi(f)|} = e^{-\Omega(i)}\]
\end{theorem}

A more explicit bound, in terms of the number of arguments and the
arity of the connective, is possible.  We use a more explicit version
of Lemma~\ref{lem:restrict} and bound the convergence of the growth
processes characterized by Theorem 5.3 in~\cite{Sa90}.   A corollary of
Lemma~\ref{lem:bound_rl} is that the same bound also applies to the
growth processes on monotone formulas whose limiting distribution is
uniform over the slice functions.

\begin{lemma}\label{LEM:BOUND_RL}\label{lem:bound_rl}
If $\alpha$ is a balanced monotone connective, then for all $w \in \cF_n$,
\[\Delta_i(w)=O(\epsilon^{2^i})+(-1)^\iprod{\upsh}{w}\Delta_i(w \wedge \chih).\]
\end{lemma}
\begin{proof}
See appendix.
\end{proof}

\begin{theorem}[(Savick\'y, 5.3 in \cite{Sa90})]\label{thm:t53}
Let $\zerobf < w \in \cF_n$, let $\alpha$ be a $k$-adic nonlinear
balanced connective, and assume that the initial distribution is
uniform over the projections, negations, and constants.  The
$\lim_{i\rightarrow\infty} \Delta_i(w) = 0$.
\end{theorem}

\begin{lemma}\label{LEM:W3P_EQ}\label{lem:w3p_eq}
Assume that the conditions of Theorem~\ref{thm:t53} are satisfied and
let $a = \sum_{t\in B_k}|S_\alpha(t)|^3 < 1 - 2^{-k}$.  If $|w| = d
\geq 2$ and
\begin{equation}\label{equ:id}
i_d = n2^k\log(a^{-1}) + \sum_{j=3}^d \frac{(k+1)^j j}{\log(a^{-1})},
\end{equation}
then $|\Delta_i(w)| \leq a^{i - i_d} b_d(i)$, where $b_d(i) =
(i - i_2 + 2)^{(k+1)^{d-3}}$, and $b_2(i) = 1$.
\end{lemma}
\begin{proof}
See appendix.
\end{proof}

\begin{theorem}\label{thm:savbounds}
Assume that the conditions of Theorem~\ref{thm:t53} are satisfied, let 
$a$ be as in Lemma~\ref{lem:w3p_eq}, and let
\[I = n2^k\log(a^{-1}) + \frac{2^{2n}(k+1)^{2^n}}{\log(a^{-1})}.\]
For any positive $c < 1$, if $w \not= \zerobf$ and 
\begin{equation}\label{eqn:final}
i \geq \frac{\log(c)}{\log(a)} +
          \frac{\log(i - I + 2)}{\log(a^{-1})}(k+1)^{2^n} + I,
\end{equation}
then $|\Delta_i(w)| \leq c$.
\end{theorem}
\begin{proof}
By Lemma~\ref{lem:w3p_eq} the coefficient of weight $2^n$ has the greatest
converging bound:
\[|\Delta_{i+1}(w)| \leq a^{i-i_{2^n}}(i - n2^k\log(a^{-1}) + 2)^{(k+1)^{2^n}}\]
where
\begin{eqnarray*}
i_{2^n} & = & n2^k\log(a^{-1}) +\sum_{j=3}^{2^n}\frac{(k+1)^j j}{\log(a^{-1})}\\
        & \leq & n2^k\log(a^{-1}) + \frac{2^{2n}(k+1)^{2^n}}{\log(a^{-1})} = I
\end{eqnarray*}
Solving for $i$ in the inequality
\[a^{i-i_{2^n}}(i - n2^k\log(a^{-1}) + 2)^{(k+1)^{2^n}} < c\]
completes the proof.
\end{proof}

Thus, by equation~\ref{eqn:inv}
\begin{eqnarray*}
\pi_i(g) 
  &=& \frac{1}{2^{2^n}}\sum_{f \in \cF_n} (-1)^\iprod{f}{g} \Delta_i(f)\\
  &=& \frac{1}{2^{2^n}}+\frac{1}{2^{2^n}}\sum_{f\in\cF_n\backslash\zerobf}
                  (-1)^\iprod{f}{g} \Delta_i(f)\\
  &\leq& \frac{1}{2^{2^n}} +
         \frac{1}{2^{2^n}}\sum_{f \in \cF_n\backslash \zerobf} |\Delta_i(f)|\\
  &\leq& \frac{1}{2^{2^n}}+\max_{f\in\cF_n\backslash\zerobf}|\Delta_i(f)|,
\end{eqnarray*}
implying that for all $g \in \cF_n$, $|\pi(g) - \pi_i(g)| \leq c$ if $i$ 
satisfies equation~\ref{eqn:final}.

\subsubsection{Varying the Initial Distribution}
Theorem~\ref{thm:result} can easily be modified to cover the cases in
which one of the constants is missing from the support of $\mu$:  in
these cases there is concentration on a single function when $n$ is
even and uniform distribution on a set of slice functions when $n$ is
odd.  When the support of $\mu$ consists only of the projection
functions, however, the situation can be more complicated.  If $\alpha$
is not self-dual or $n$ is odd, the result is the same as when both
constants are present.  If $\alpha$ is self-dual and $n$ is even,
however, the result is given by the following theorem.

\begin{theorem}
Let $\alpha$ be a monotone self-dual non-projection connective, $n$ be
even and the support of $\mu$ comprise the projection functions.  Then
the limiting distribution is uniform on the self-dual functions in
$\cS_{n/2,n}$.
\end{theorem}
\begin{proof}
Same as theorem~\ref{thm:self_dual}.
\end{proof}

We note that there are $2^{\Half{n\choose n/2}}$ self-dual functions in
$\cS_{n/2,n}$.  According to a result of Sapozhenko~\cite{Sa89}, this
is only a tiny fraction, less than $\exp -({n\choose n/2+1}
2^{-n/2-1})$, of the support of the growth process, which is the set of
self-dual monotone functions.

\section{Growing Other Functions}\label{sec:other}
We can use the same method to analyze other growth processes.  For
example, the uniform distribution on the set of bi-preserving functions
(that is, those functions satisfying $f(0, \ldots, 0) = 0$ and $f(1,
\ldots, 1) = 1$) can be generated by a growth process that uses the
bi-preserving selection connective $\alpha(x,y,z) = xy \vee \bar{x}z$
and an initial distribution that is uniform on the projection
functions.  The same technique as in the monotone case is sufficient to
prove this; the corresponding restriction lemma yields the identity
\[\lim_{i\rightarrow\infty}\Delta_i(w) = (-1)^\iprod{w}{\eta_n}
                                        \Delta(w \wedge \kappa_n)\]
where $\eta_n = \wedge_{j=1}^n x_j$ and $\kappa_n = \vee_{j=1}^n x_j
\bigwedge \overline{\wedge_{j=1}^n x_j}$.  Similar analysis for the
$0$-preserving and $1$-preserving functions follows easily.

\section{Conclusion}\label{sec:conc}
In this paper we have developed techniques for analyzing growth processes
when the limiting distribution is uniform over a set of Boolean functions.
In particular, we can deal with situations in which this set comprises
neither a single function nor all Boolean functions.

We believe that straightforward extensions of the techniques used here
will yield a classification of all situations leading to uniform
distributions over sets of functions.  The step that remains to be
taken is the classification of all sets of functions that can be
computed by formulas that are complete $k$-ary trees built from a
single connective.  There is some work by Kudryavtsev~\cite{Ku60,Ku60b}
on this problem, but it stops short of a complete classification.

A more adventurous direction for further work is to deal with
situations leading to non-uniform distributions.  Empirical
computations show that these situations can be quite complicated, and
we are not yet able to formulate a conjecture that covers all our
data.

\bibliography{rcf}

\appendix

\section{Proofs of Lemmas \ref{lem:fast-nfp}, \ref{lem:slow-nfp}, 
                          \ref{lem:triv}, \ref{lem:minslope}, and
                          \ref{lem:bound_rl}}
\begin{fact} If $\alpha$ is a monotone connective, then:
\begin{enumerate}
\setlength{\parskip}{0pt}
\setlength{\partopsep}{0pt}
\setlength{\parsep}{0pt}
\setlength{\itemsep}{0pt}
\item $A_\alpha(p) > p$ on the interval $(0,1)$ if and only if 
      $\alpha(x) = x_i \vee \alphap(x)$ if and only if $\beta_1 > 0$.
\item $\beta_{k-1} = \frac{a}{k},\ a \in \{0,1,\ldots,k-1\}$.
\item $\beta_0 = 0$ and $\beta_k = 1$.
\item $A_\alpha(s) = s \in (0,1)$ if and only if $A_\alpha(p) < p,\ p 
      \in (0,s)$ and $A_\alpha(p) > p,\ p \in (s,1)$ if and only if
      $\Ap_\alpha(0) = \Ap_\alpha(1) = 0$ if and only if $\beta_0 = \beta_1 = 0$
      and $\beta_k = \beta_{k-1} = 1$.
\end{enumerate}
\end{fact}

\subsection{Proof of Lemma~\ref{lem:fast-nfp}}
\begin{proof}
Begin by differentiating $A_\alpha(p)$
\begin{eqnarray*}
\Ap_\alpha(p) 
  & = & \frac{d}{dp}\left(\sum_{i=0}^k \beta_i\Ch{k}{i}p^i(1-p)^{k-i}\right) \\
  & = & \frac{d}{dp}\left( p^k + \beta_{k-1}kp^{k-1}(1-p) + 
                       \sum_{i=0}^{k-2}\beta_i\Ch{k}{i}p^i(1-p)^{k-i} \right)\\
  & = & kp^{k-1} + \beta_{k-1}kp^{k-2}(k-kp-1) 
             - \sum_{i=0}^{k-2}\beta_i\Ch{k}{i}p^{i-1}(1-p)^{k-i-1}(pk-i),
\end{eqnarray*}
and evaluate at $1-\epsilon$ 
\begin{eqnarray*}
\Ap_\alpha(1-\epsilon) 
   & = & k(1-\epsilon)^{k-2}(1-\epsilon + \beta_{k-1}k\epsilon-\beta_{k-1}) 
          - \sum_{i=0}^{k-2}\beta_i\Ch{k}{i}(1-\epsilon)^{i-1}
         (\epsilon)^{k-i-1}(k-i-k\epsilon) \\
   & > & k(1-\epsilon)^{k-2}(1-\epsilon + \beta_{k-1}k\epsilon-\beta_{k-1}) 
          - \sum_{i=0}^{k-2}\beta_i\Ch{k}{i}(1-\epsilon)^{i-1}
         (\epsilon)^{k-i-1}k \\
   & > & k(1-\epsilon)^{k-2}(1-\epsilon + \beta_{k-1}k\epsilon-\beta_{k-1}) 
          - \sum_{i=0}^{k-2}\beta_i\Ch{k}{i} (\epsilon)^{k-i-1}k  \\
   & = & k(1-\epsilon)^{k-2}(1-\epsilon + \beta_{k-1}k\epsilon-\beta_{k-1}) 
          - k\epsilon \sum_{i=0}^{k-2}\beta_i\Ch{k}{i} (\epsilon)^{k-i-2} \\
   & > & k(1-\epsilon)^{k-2}(1-\epsilon + \beta_{k-1}k\epsilon-\beta_{k-1}) 
          - k\epsilon \sum_{i=0}^k\Ch{k}{i} \\
   & > & k(1-\epsilon)^{k-2}(1-\epsilon + \frac{k-2}{k}(k\epsilon-1)) 
         - k\epsilon 2^k \\
   & = & (1-\epsilon)^{k-2}(2 + k\epsilon(k - 3)) - k\epsilon 2^k.\\
\end{eqnarray*}
For $k > 2$ and $\epsilon < \epsilon_k$, 
\begin{eqnarray*}
\Ap_\alpha(1-\epsilon) 
   & = & (1-\epsilon)^{k-2}(2 + k\epsilon(k - 3)) - k\epsilon 2^k \\
   & > & (1-\epsilon)^{k-2}(2 + k\epsilon(k - 3)) - \Half \\
   & > & (1-\frac{1}{48})2  - \Half \\
   & = & \frac{35}{24} \\
\end{eqnarray*}
\end{proof}

\subsection{Proof of Lemma~\ref{lem:slow-nfp}}
\begin{proof} 
For the lower bound, observe that since
\[\Ap_\alpha(p) = kp^{k-1} - 
                  \sum_{i=0}^{k-1}\beta_i\Ch{k}{i}p^{i-1}(1-p)^{k-i-1}(pk-i),\]
for all $p > 1 - k^{-1}$, minimizing $\Ap_\alpha(p)$, maximizes the
coefficients $\beta_i$, for all $i=2\ldots k-2$.  Since the connective is
of the form $\alpha(x) = x_j \wedge \alphap(x)$, $\beta_i \leq
\Ch{k-1}{i}\Ch{k}{i}^{-1} = \frac{k-i}{k}$.  Hence,
\[\Ap_\alpha(p) 
   \geq kp^{k-1}-\sum_{i=0}^{k-1}\Ch{k-1}{i}p^{i-1}(1-p)^{k-i-1}(pk-i)
      = 1 + (1-p)^{k-2}(kp - 1).\]
Thus, for all $\epsilon < k^{-1}$
\[\Ap_\alpha(1-\epsilon) \geq 1 + \epsilon^{k-2}(k-1-k\epsilon) 
                            > 1 + \epsilon^{k-2}  
                            > 1 + \epsilon^k.\]

For the upper bound we minimize all $\beta_i$ for $i = 2\ldots k-2$, i.e., 
$\beta_i = 0$.  Thus, for all $p > 1 - k^{-1}$,
\[\Ap_\alpha(p) \leq kp^{k-1} - (k-1)p^{k-2}(pk-k+1) = p^{k-2}(k(k-2)(1-p)+1),\]
implying that for all $\epsilon < k^{-1}$,
\[\Ap_\alpha(1-\epsilon) \leq (1-\epsilon)^{k-2}(k(k-2)\epsilon + 1).\]
\end{proof}

\subsection{Proof of Lemma~\ref{lem:triv}}
\begin{proof}
Since $\alpha$ is not of the form $\alpha(x) = x_i \vee \alphap(x)$, the 
first two coefficient of $A_\alpha$ are $\beta_0 = \beta_1 = 0$.  Thus, on 
the interval $(0,1)$,
\[A_\alpha(p) = \beta_2\Ch{k}{2}p^2 + B(p)p^3 \leq \Ch{k}{2}p^2 + B(p)p^2 < 
         (\Ch{k}{2}+1)p^2\]
since $B(p) < 1$ on the interval $(0,1)$.
\end{proof}

\subsection{Proof of Lemma~\ref{lem:minslope}}
\begin{proof}
Consider a projection connective, say $\chi(x) = x_b$, $b \in [1,n]$.  
The corresponding characteristic polynomial is 
\[A_\chi(p) = p = p^k + p(1-p)^{k-1} + 
                  \sum_{i=2}^{k-1} \Ch{k-1}{i}p^i(1-p)^{k-i},\]
whose fixed-point is everywhere and whose slope is $1$.  Note that
$\beta_1 = 1$ and $\beta_{k-1} = \frac{k-1}{k}$.  Let 
\[\eta(x) = \wedge_{i=2}^k x_i \bigvee \vee_{i=2}^k (x_1\wedge x_i),\]
be a $k$-adic monotone connective. The corresponding characteristic
polynomial 
\begin{eqnarray*}
A_\eta(p) & = & p - p(1-p)^{k-1} + p^{k-1}(1-p) \\
          & = & A_\chi(p) - p(1-p)^{k-1} + p^{k-1}(1-p)\\
          & = & A_\chi(p) + (A_\eta - A_\chi(p))\\
\end{eqnarray*}
has a fixed-point at $\Half$.  Not surprisingly, this is almost
$A_\chi(p)$ except that $\beta_1 = 0$ and $\beta_{k-1} = 1$, i.e., the
difference is just two terms.  We claim that $\Ap_\eta(\Half) \leq
\Ap_\alpha(s)$.

The claim is proved by contradiction.  Assume that $\Ap_\eta(\Half)$ is
not the minimum slope at a fixed-point, then there exists a $k$-adic
monotone connective $\zeta$, whose degree $k$ characteristic polynomial
$A_\zeta(p)$ has a fixed-point $t \in (0,1)$, such that $\Ap_\zeta(t) <
\Ap_\eta(\Half)$.  Since $\beta_1 = 0$ and $\beta_{k-1} = 1$ must hold
for $A_\zeta$, we write $A_\zeta(p)$ in a manner similar to $A_\eta(p)$: 
$A_\zeta(p) = A_\eta(p) + (A_\zeta(p) - A_\eta(p))$.  Specifically we 
are interested in the differences between $A_\zeta(p)$ and $A_\eta(p)$.  
In fact,
\begin{eqnarray*}
A_\zeta(p) 
      & = & A_\eta(p) 
           + \left[ p^{i_0}(1-p)^{k-i_0} + p^{i_1}(1-p)^{k-i_1} \ldots \right]\\
      &  & - \left[ p^{j_0}(1-p)^{k-j_0} + p^{j_1}(1-p)^{k-j_1} \ldots \right],
\end{eqnarray*}
where $i_l \leq i_{l+1} \leq k-2$ and $j_l \geq j_{l+1} \geq 2$. 

Since $A_\zeta(p) \not= A_\eta(p)$, and $A_\zeta(p)$ has a fixed-point
on $(0,1)$, either $i_0$ or $j_0$ must exist.  Without loss of
generality assume that $i_0$ exists and consider the characteristic
polynomial $A_\delta(p) = A_\zeta(p) - p^{i_0}(1-p)^{k-i_0}$.  Since
the connective corresponding to $A_\zeta(p)$ is monotone, $j_0 < kt <
i_0$. which implies that the derivative $p^{i_0-1}(1-p)^{k-i_0-1}(i_0-kp)$ 
of the term $p^{i_0}(1-p)^{k-i_0}$ is positive for all $u \leq t$.
Since the fixed-point $u$ of $A_\delta(p)$ is in the interval
$(\frac{j_0}{k},t)$, $\Ap_\delta(u) < \Ap_\zeta(t)$, which is a
contradiction!  In fact, iteratively subtracting the terms
$p^{i_l}(1-p)^{k-i_l}$ and adding the terms $p^{j_l}(1-p)^{k-j_l}$,
reduces $A_\zeta(p)$ to $A_\eta(p)$!.

Evaluating $\Ap_\eta(\Half) = 1 + \frac{k-2}{2^{k-2}}$ completes the proof.
\end{proof}

\subsection{Proof of Lemma~\ref{lem:bound_rl}}
\begin{proof}
\newcommand{\wcxh}{{w\cap\chih}}
Following the proof Lemma~\ref{lem:restrict} begin with the definition
\[\Delta_i(w) = \sum_{v \in B_n} (-1)^\iprod{v}{w} \pi_i(v),\]
and rewrite the equation as
\[\Delta_i(w) = \sum_{v \in B_n} (-1)^\iprod{v}{w}\pi_i(v)
              = \sum_{t \leq \chih}\sum_{u \leq \notchih}
                   (-1)^\iprod{t\vee u}{w} \pi_i(t \vee u).\]
Consider the restriction of $w$ to the slice $\frac{n}{2}$, that is,
$w \wedge \chih$.  By Theorem~\ref{thm:bound-fp}, after $O(\log(n))$ iterations
$\pi_i(t \vee u) = O(\epsilon^{2^i})$, if $u \not= \upsh$. Hence, for $i >
O(\log(n))$, $\Delta_i(w\wedge \chih)$ can be rewritten as

\begin{eqnarray*}
\Delta_i(w\wedge \chih)
 &=& \sum_{v \in B_n} (-1)^\iprod{v}{w\wedge \chih}\pi_i(v) \\
 &=& \sum_{t\leq\chih}\sum_{u \leq \notchih}
                   (-1)^\iprod{t\vee u}{w\wedge \chih} \pi_i(t \vee u) \\
 &=& \sum_{t\leq\chih}
                   (-1)^\iprod{t\vee \upsh}{w\wedge \chih} \pi_i(t \vee \upsh) +
     \sum_{t\leq\chih}\sum_{u \leq \notchih, u\not= \upsh}
                   (-1)^\iprod{t\vee u}{w\wedge \chih} \pi_i(t \vee u)\\
 &=& \sum_{t\leq\chih}
                   (-1)^\iprod{t\vee \upsh}{w\wedge \chih} \pi_i(t \vee \upsh) +
     \sum_{t\leq\chih}\sum_{u \leq \notchih, u\not= \upsh} O(\epsilon^{2^i})\\
% &=&O(\epsilon^{2^i}) +
%  \sum_{t\leq\chih}(-1)^\iprod{t\vee \upsh}{w\wedge \chih}\pi_i(t \vee \upsh)\\
% &=& O(\epsilon^{2^i}) + \sum_{t \leq \chih} (-1)^\iprod{t}{w\wedge \chih}
%                        (-1)^\iprod{\upsh}{w\wedge \chih} \pi_i(t \vee \upsh)\\
% &=& O(\epsilon^{2^i}) +
%    \sum_{t \leq \chih} (-1)^\iprod{t}{w\wedge \chih} \pi_i(t \vee \upsh) \\
 &=& O(\epsilon^{2^i}) +
     \sum_{t \leq \chih} (-1)^\iprod{t}{w} \pi_i(t \vee \upsh).
\end{eqnarray*}
This, in conjunction with
\begin{eqnarray*}
\Delta_i(w)
 &=& \sum_{t\leq\chih}\sum_{u \leq \notchih}
                   (-1)^\iprod{t\vee u}{w} \pi_i(t \vee u) \\
 &=& \sum_{t\leq\chih} (-1)^\iprod{t\vee \upsh}{w} \pi_i(t \vee \upsh) +
     \sum_{t\leq\chih}\sum_{u \leq \notchih, u\not= \upsh}
                   (-1)^\iprod{t\vee u}{w} \pi_i(t \vee u) \\
 &=& \sum_{t\leq\chih} (-1)^\iprod{t\vee \upsh}{w} \pi_i(t \vee \upsh) +
     \sum_{t\leq\chih}\sum_{u \leq \notchih, u\not= \upsh} O(\epsilon^{2^i})\\
% &=& O(\epsilon^{2^i}) +
%     \sum_{t \leq \chih} (-1)^\iprod{t \vee \upsh}{w} \pi_i(t \vee \upsh)\\
% &=& O(\epsilon^{2^i}) +
%     \sum_{t \leq \chih} (-1)^\iprod{t}{w}(-1)^\iprod{\upsh}{w}
%                                 \pi_i(t \vee \upsh) \\
 &=& O(\epsilon^{2^i}) +
     (-1)^\iprod{\upsh}{w} \sum_{t \leq \chih} (-1)^\iprod{t}{w}
                                 \pi_i(t \vee \upsh) \\
 &=& O(\epsilon^{2^i}) + (-1)^\iprod{\upsh}{w} \Delta_i(w\wedge \chih)
\end{eqnarray*}
yields the identity.
\end{proof}

\section{Proof of Lemma {\ref{lem:w3p_eq}}}
\begin{lemma}[(Savick\'y, 1990, Lemma 5.1 in~\cite{Sa90})]\label{lem:beq}\ \\
Let $x_{i+1} = \sum_{j=1}^k a_jx_i^j$, such that $|x_0| < 1$,
$\sum_{j=1}^k |a_j| \leq 1$, and $|a_1| < 1$.  Then $|x_{i+1}| \leq
a|x_i|$, where $a \leq |a_1| + |x_0|(1-|a_1|)$.  Hence,
$|x_i| \leq a^i|x_0|$.
\end{lemma}

\begin{lemma}[(Savick\'y, 1990, Lemma 4.7 in~\cite{Sa90})]\ \\
For any $k$-adic connective $\alpha$, $\sum_{t \in B_k} S_\alpha(1)^2 = 1$, 
and $\alpha$ is balanced and nonlinear if and only if $S_\alpha(\zerobf) = 0$
and for all $s \in B_k$, $|S_\alpha(s)| < 1$.
\end{lemma}

\begin{lemma}\label{lem:w2_eq}
Let $\alpha$ be a balanced nonlinear $k$-adic connective and let $|w| = 2$.  
For any positive $c < 1$, if 
\[i > \log(c^{-1})n2^k,\]
then $|\Delta_i(w)| \leq c$.
\end{lemma}
\begin{proof}
By Theorem 5.3 in~\cite{Sa90}, 
\[\Delta_i(w) = \sum_{j=1}^k a_j(w) \Delta_{i-1}(w)^j,\]
where
\[a_j(w)\ =\ \sum_\Sarg{t \in B_k}{|t|=j} S_\alpha(t)^{|w|} 
        \ =\ \sum_\Sarg{t \in B_k}{|t|=j} S_\alpha(t)^2.\]
By corollary~\ref{lem:beq}, $|\Delta_i(w)| \leq a^i |\Delta_0(w)|$,
where $a \leq |a_1(w) + \Delta_0(w)(1-a_1(w))$.  By
Theorem 5.3 in~\cite{Sa90}, $\Delta_0(w) < \frac{n-1}{n+1}$, hence,

\[a\ \leq\ |a_1(w) + \frac{n-1}{n+1}(1-a_1(w))\ =\ 1-\frac{2}{n+1}(1-a_1(w)),\]
and since $a_1(w) \leq 1-2^{-k}$, 
\[a \leq 1 - \frac{2}{n+1}2^{-k}.\]

Solving for $i$ in the inequality $a^i|\Delta_0(w)| \leq c$ yields:
\[i \geq \frac{\log(c) - \log|x_0|}{\log(a)},\]
and substituting for $a$ on the right:
\begin{eqnarray*}
\frac{\log(c) - \log|x_0|}{\log(a)} 
& = & \frac{\log(c) - \log|x_0|}{\log\left(1 - \frac{2}{n+1}2^{-k}\right)} \\
& < & ((n+1)2^{k-1} + \frac{1}{2})(\log(c^{-1}) + \log|x_0|) \\
& < & ((n+1)2^{k-1} + \frac{1}{2})\log(c^{-1}) \\
& < & n2^k\log(c^{-1}) \\
\end{eqnarray*}

Thus, for $i > \log(c^{-1})n2^k$,\ $|\Delta_i(w)| \leq c$.
\end{proof}

\begin{lemma}[(Savick\'y, 1990, Lemma 5.2 in~\cite{Sa90})]\label{lem:feq}\ \\
Let $x_{i+1} = y_i + \sum_{j=1}^k a_jx_i^j$, such that $|x_i| \leq 1$,
and $a = \sum_{j=1}^k |a_j| < 1$.  Then $|x_{i+1}| \leq a|x_i| + y_i$.
\end{lemma}

\begin{corollary}\label{cor:nl}
If $y_i < (i-l+2)^k a^{i-l+1}< a$ for some $k \geq 0$ and $l > 0$, then 
$|x_i| \leq (i-l+1)^{k+1}a^{i-l}$ for all $i \geq l$.
\end{corollary}
\begin{proof}
By Lemma~\ref{lem:feq}, $|x_{i+1}| \leq a|x_i| + y_i$.  Hence, by 
induction on $i$
\begin{eqnarray*}
|x_{i+l}| & \leq & a^i(|x_l| + (i-l+2)^ka^l) \\
          & \leq & a^{i-l}\left(1 + \sum_{j=l}^i(i-l+2)^k\right) \\
          & \leq & a^{i-l}(i-l+2)^{k+1}
\end{eqnarray*}
\end{proof}

\subsection{Proof of Lemma~\ref{lem:w3p_eq}}
\begin{proof}
The proof is by induction on $j$.  By Theorem 5.3 in~\cite{Sa90}, 
\[\Delta_i(w)\ =\ y_i(w) + \sum_{j=1}^k a_j(w) \Delta_{i-1}(w)^j ,\]
where
\[a_j(w) = \sum_{\{t \in B_k:\ |t|=j\}} S_\alpha(t)^{|w|},
         = \sum_{\{t \in B_k:\ |t|=j\}} S_\alpha(t)^d,\]
and
\[y_i(w) = \sum_{\vbf \in G_w^k} 
                 \left(\prod_{a\in B_n| w(a) = 1} S_\alpha(\vbf(a))\right)
                 \left(\prod_{j = 1}^k \Delta_i(\vbf_j)\right),\]
where $G_w = \{v \in B_{2^n}\ |\ \zerobf < v < w\}$.

Since $\sum_{t \in B_k} S_\alpha(1)^2 = 1$ and $|S_\alpha(s)| < 1$ for all
$s \in B_k$, 
\begin{eqnarray*}
a(w) &  =   & \sum_{j=1}^k a_j(w) 
     \  =   \ \sum_{j=1}^k \sum_{t \in B_k: |t| = j} S_\alpha(t)^{|w|} \\
     &  =   & \sum_{t \in B_k} S_\alpha(t)^{|w|}
     \ \leq \ \sum_{t \in B_k} |S_\alpha(t)|^{|w|} \\
     & \leq & \sum_{t \in B_k} |S_\alpha(t)|^3 
     \  <   \ \sum_{t \in B_k} S_\alpha(t)^2 =  1,
\end{eqnarray*}
The base case, $d = 2$, is proved by Lemma~\ref{lem:w2_eq}.  For the
base case, $d = 3$, we first bound $y_i(w)$.  By Theorem 5.3
in~\cite{Sa90}, $\Delta_i(w) = 0$ if $|w| = 1$ and by
Lemma~\ref{lem:w2_eq}, $\Delta_i(w) \leq a^{i - n2^k\log(a^{-1})}$ if
$|w| = 2$.  Hence, for all $w \in G_w$, then $\Delta_i(w) \leq a^{i -
n2^k\log(a^{-1})}$.  If we let $i_2 = n2^k\log(a^{-1})$ then
$\Delta_i(w) \leq a^{i - i_2}b_2(i)$ and,
\begin{eqnarray*}
y_i(w) & = & \sum_{\vbf \in G_w^k} 
                 \left(\prod_{a\in B_n| w(a) = 1} S_\alpha(\vbf(a))\right)
                 \left(\prod_{j = 1}^k \Delta_i(\vbf_j)\right) \\
   & < & \sum_{\vbf \in G_w^k}\prod_{j = 1}^k \Delta_i(\vbf_j) \\
   & < & \sum_{\vbf \in G_w^k} \left(a^{i-i_{d-1}}b_{d-1}(i)\right)^k \\
   & = & \left(\sum_{j=1}^{d-1}\Ch{d}{j}\right)^k
               \left(a^{i-i_{d-1}}b_{d-1}(i)\right)^k \\
   & < & \left(2^d a^{i-i_{d-1}}b_{d-1}(i)\right)^k.
\end{eqnarray*}
Solving for $i$ the inequality 
\[\left(2^d a^{i-i_{d-1}}b_{d-1}(i)\right)^k\ <\ a\]
yields
\[\frac{d + \log(b_{d-1}(i))}{\log(a^{-1})} + k^{-1} + i_{d-1}\ \leq\ 
  \frac{(k+1)^dd}{\log(a^{-1})} + i_{d-1}\ =\ i_d\ \leq\  i,\]
which equals to equation~\ref{equ:id} when evaluated at $d=3$. 
Hence, for $i \geq i_d$
\[y_i(w)\ <\ \left(2^d a^{i-i_{d-1}}b_{d-1}(i)\right)^k  
        \ <\  b_{d-1}(i)^k a^{i-i_d} < a,\]
and by Lemma~\ref{lem:feq} and Corollary~\ref{cor:nl} we complete the 
base case:
\begin{eqnarray*}
|\Delta_{i+1}(w)| & \leq & a|\Delta_i(w)| + y_i(w)  \\
                  & \leq & a|\Delta_i(w)| + a^{i-i_d}b_{d-1}(i)^k \\
                  & \leq & a^{i-i_d}\left(1 + \sum_{j=i_d}^i 1\right) \\
                  & \leq & a^{i-i_d}(i-i_d+2) \\
                  & \leq & a^{i-i_d}(i-i_2+2) \\
                  & =    & a^{i-i_d}b_d(i).
\end{eqnarray*}

Assume that the hypothesis holds for all $w$ of weight less than some 
fixed $d$ and let $|w| = d$.  Repeating the above calculations in terms 
of $d$ we get an identical bound for $y_i(w)$
\[y_i(w)\ <\ \left(2^d a^{i-i_{d-1}}b_{d-1}(i)\right)^k  
        \ <\  b_{d-1}(i)^k a^{i-i_d} < a,\]
where, by the inductive hypothesis,
\begin{eqnarray*}
i_d & = & \frac{(k+1)^dd}{\log(a^{-1})} + i_{d-1} \\
    & = & \frac{(k+1)^dd}{\log(a^{-1})} + n2^k\log(a^{-1}) + 
              \sum_{j=3}^{d-1} \frac{(k+1)^j j}{\log(a^{-1})}\\
    & = & n2^k\log(a^{-1}) + \sum_{j=3}^d \frac{(k+1)^j j}{\log(a^{-1})}
\end{eqnarray*}
and hence
\begin{eqnarray*}
|\Delta_{i+1}(w)| & \leq & a|\Delta_i(w)| + y_i(w)  \\
      & \leq & a|\Delta_i(w)| + a^{i-i_d}b_{d-1}(i)^k \\
      & \leq & a^{i-i_d}\left(1 + \sum_{j=i_d}^i b_{d-1}^k\right) \\
      & \leq & a^{i-i_d}\left(1 + \sum_{j=i_d}^i 
                                  (i-i_2+2)^{(k+1)^{d-1-3}k}\right)\\
      & \leq & a^{i-i_d} (i-i_d+1) (i-i_2+2)^{(k+1)^{d-1-3}k} \\
      & \leq & a^{i-i_d} (i-i_2+2) (i-i_2+2)^{(k+1)^{d-1-3}k} \\
      & =    & a^{i-i_d} (i-i_2+2)^{(k+1)^{d-3}} \\
      & =    & a^{i-i_d}b_d(i).
\end{eqnarray*}
completing inductive step.
\end{proof}

\end{document}